\documentclass[fleqn,usenatbib]{mnras}

\usepackage{newtxtext,newtxmath}
\usepackage{color}

\usepackage[T1]{fontenc}
\usepackage{ae,aecompl}

\hypersetup{draft}

\usepackage{graphicx}	% Including figure files
\usepackage{amsmath}	% Advanced maths commands
\usepackage{amssymb}	% Extra maths symbols

\title[Monte-Carlo simulations of black hole mergers in AGN disks]{Monte-Carlo simulations of black hole mergers in AGN disks: Low $\chi_{\rm eff}$ mergers and predictions for LIGO}

\author[B. McKernan et al.]{B. McKernan$^{1,2,3}$\thanks{E-mail:bmckernan at amnh.org (BMcK)}, K.E.S. Ford$^{1,2,3}$, R. O'Shaugnessy$^{4}$ \& D. Wysocki$^{4}$\\
$^{1}$Department of Astrophysics, American Museum of Natural History, New York, NY 10024, USA\\
$^{2}$Graduate Center, City University of New York, 365 5th Avenue, New York, NY 10016, USA\\
$^{3}$Department of Science, BMCC, City University of New York, New York, NY 10007, USA\\
$^{4}$Center for Computational Relativity and Gravitation, Rochester Institute of Technology, Rochester, NY 14623, USA\\
}

\date{Accepted XXX. Received YYY; in original form ZZZ}

\pubyear{2019}

\begin{document}
\label{firstpage}
\pagerange{\pageref{firstpage}--\pageref{lastpage}}
\maketitle

\begin{abstract}
Accretion disks around supermassive black holes are promising sites for stellar mass black hole mergers detectable with LIGO. Here we present the results of Monte-Carlo simulations of black hole mergers within 1-d AGN disk models. For the spin distribution in the disk bulk, key findings are: (1) The distribution of $\chi_{\rm eff}$ is naturally centered around $\tilde{\chi}_{\rm eff} \approx 0.0$, (2) the width of the $\chi_{\rm eff}$ distribution is  narrow for low natal spins. For the mass distribution in the disk bulk, key findings are: (3) mass ratios  $\tilde{q} \sim 0.5-0.7$, (4) the maximum merger mass in the bulk is $\sim 100-200M_{\odot}$, (5) $\sim 1\%$ of bulk mergers involve BH $>50M_{\odot}$ with (6) $\simeq 80\%$ of bulk mergers are pairs of 1st generation BH. Additionally, mergers at a migration trap grow an IMBH with typical merger mass ratios $\tilde{q}\sim 0.1$. 
Ongoing LIGO non-detections of black holes $>10^{2}M_{\odot}$ puts strong limits on the presence of migration traps in AGN disks (and therefore AGN disk density and structure) as well as median AGN disk lifetime. The highest merger rate occurs for this channel if AGN disks are relatively short-lived ($\leq 1$Myr) so multiple AGN episodes can happen per Galactic nucleus in a Hubble time.
\end{abstract}


%%%%%%%%%%%%%%%%%%%%%%%%%%%%%%%
\section{Introduction}
Advanced LIGO \citep{Aasi15} and Advanced Virgo \citep{Acernese15} have revealed a population of merging black holes (BHs) that is surprisingly numerous at the upper-end of the rate estimate, and  significantly more massive than those observed in our own Galaxy \citep{LIGO18}. Most tantalizingly, the first few mergers are associated with a low observed $\chi_{\rm eff}$, which could imply a precursor population of BHs biased towards low spin, or a precursor population of BHs biased towards anti-alignment or a wide range of spin alignments. Important channels for BH merger are likely to include field binary mergers \citep[e.g.][]{Belczynski10,DeMink16}, dynamical mergers in globular clusters or galactic nuclei \citep[e.g.][]{Rodriguez16,Antonini16} and AGN disks \citep{McK14, Yang19}. The next few years of LIGO operation will increase the population of black hole mergers and allow us to construct underlying prior distributions of mass and spin in the precursor BHs \citep[e.g.][]{Fishbach17,Gerosa17,Wysocki18}.\\

In the local Universe, black hole density seems greatest in our own Galactic nucleus. The observed rate of occurrence of black hole X-ray binaries implies a cusp of BHs in the central parsec \citep{Hailey18,Generozov18} consistent with previous conjectures \citep{Morris93,Miralda00}. As a result one of the most potentially promising sites to generate a high rate of black hole mergers yielding over-massive BHs detectable with LIGO are AGN disks \citep{McK12,McK14,Bartos17,Stone17,McK17,Secunda18}. The hints from LIGO/Virgo's O1 and O2 of low $\chi_{\rm eff}$ are intriguing and hopefully strongly constraining -- we require not only massive BH's, but also binary spin configurations that produce low $\chi_{\mathrm{eff}}$ often enough to explain the mergers in O1 and O2.

\section{Low $\chi_{\rm eff}$ does not necessarily mean low spin}
To provide context for our simulation results, let us consider how low $\chi_{\rm eff}$ BHs can arise in the context of AGN disk mergers.
First, it is important to understand precisely what LIGO measures when it comes to black hole spin.  Gravitational wave measurements reliably constrain the combination  \citep{Abbott16}
%\dwy{I recommend using the convention usually used in GW literature of lower-case component masses, $m_1$ and $m_2$, and upper-case used only for the total mass, $M = m_1 + m_2$.  Also, $\vec{x}$ for denoting vectors usually isn't very good typographically, and $\boldsymbol{x}$ is cleaner, though that's probably up to the journal.}
\begin{equation}
\chi_{\mathrm{eff}} = \frac {M_{1}\vec{a}_{1} +M_{2}\vec{a}_{2}}{M_{1}+M_{2}} \cdot \vec{L}
\end{equation}
when a merger occurs between two BHs of masses $M_{1},M_{2}$ and spins $\vec{a}_{1},\vec{a}_{2}$ respectively.  Notably, this combination is nearly conserved in GR and only involves the  projection of the spins onto the angular momentum ($\vec{L}$) of the merging binary. For our purposes we can rewrite $\chi_{\rm eff}$ as
\begin{equation}
\chi_{\mathrm{eff}} = \frac {M_{1}}{M_{b}}|a_{1}|\cos\phi_{1} +\frac{M_{2}}{M_{b}}|a_{2}|\cos\phi_{2}
\label{eq:chi}
\end{equation}
where ($\phi_{1},\phi_{2}$) are the angles beween the spins $\vec{a_{1}},\vec{a_{2}}$ and the orbital angular momentum of the binary $\vec{L}_{\mathrm{bin}}$, which we assume is aligned or anti-aligned with the angular momentum of the disk gas, and $M_{b}=M_{1}+M_{2}$ is the binary mass.

From eqn.~(\ref{eq:chi}), low $\chi_{\rm eff}$ could occur in a given merger if both spin magnitudes $|a_{1},a_{2}|$ are small. However, $\chi_{\rm eff}$ could also be small if spins $\vec{a}_{1},\vec{a}_{2}$ are anti-aligned so that cos($\phi_{1},\phi_{2}$) are opposite sign. Or, $\chi_{\rm eff}$ could be small if $a_{2}$ is large, $a_{1}$ is small and $m_1 \gg m_2$. Or, if both cos($\phi_{1},\phi_{2}$) are small, which can happen if both spins are oriented at large angles to $\vec{L}_{\rm bin}$, then low $\chi_{\rm eff}$ can occur.

\subsection{How we might get a $\chi_{\rm eff}$ black hole binary?}
\label{sec:low_chi}
\begin{figure}
\begin{center}
\includegraphics[width=\columnwidth]{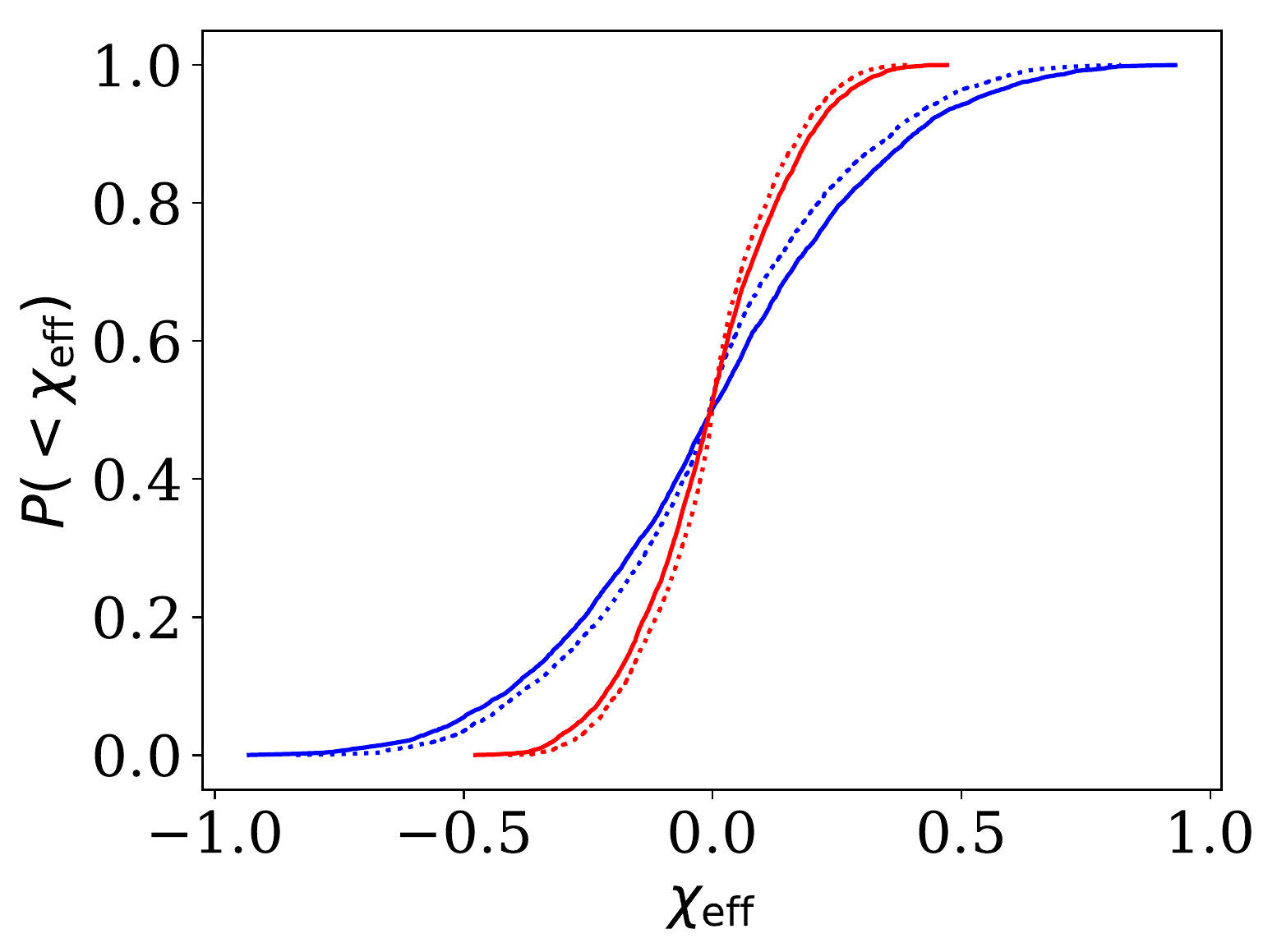}
\end{center}
\caption[acostheta vs theta]{Inferred distribution of $\chi_{\rm eff}$ for individual BHs Allowed envelope for $\chi_{\rm eff}$ for individual BHs embedded in an AGN disk drawn from a population of uniform flat spin $a\in[-1,+1]$ where the spin vectors  are oriented at angle $\phi$ compared to the disk gas. The first generation of mergers in this channel will have $\chi_{\rm eff}$ drawn from this distribution. The angle $\phi$ is drawn from a uniform distribution $\phi\in[0,\pi/2]$rad for $a>0$ and $\phi\in[\pi/2,\pi]$rad for $a<0$. Blue and red solid lines correspond to the distribution of $\chi_{\rm eff}$ under the assumptions that $a_i<1$ and $a_i<0.5$, respectively.  Dotted lines are calculated assuming $a_2=0$ (i.e., that the first spin dominates $\chi_{\rm eff}$).
\label{fig:chieff}}
\end{figure}

In this section we describe our prior assumptions for BH in AGN disks, and how these assumptions produce binaries with preferentially low $\chi_{\rm eff}$.  For BH spin magnitudes, we assume an initial  uniform spin magnitude distribution between $a=[-1,+1]$.   For spin alignment relative to the AGN gas disk, we assume the initial angle $\phi$ between a given BHs' spin and the angular momentum of the disk gas ($\vec{L}_{\rm disk}$)  is  drawn from a uniform distribution over $\phi\in[0,\pi]$, with $\phi=0(\pi)$ corresponding 
%Thus $\phi=[0,\pi/2]$rad for prograde spin BHs ($a>0$) and $\phi=[\pi/2,\pi]$rad for retrograde spin BHs ($a<0$), and $\phi=0(\pi)$rad corresponds
to perfect alignment(anti-alignment) with the disk gas angular momentum.  [This assumption modestly favors alignment or antialignment relative to the gas disk, compared to isotropic spin misalignment on the sphere.]
Third, we draw masses randomly from our population, which for the purposes of this section is uniform in $m_1,m_2$, 
%To be concrete, we have adopted a uniform mass distribution $m_1,m_2$, 
though the result does not change appreciably for a powerlaw mass distribution. % \dwy{Note that we've done this both for uniform and powerlaw, and neither changes result measurably}.  
Based on these assumptions, Fig.~\ref{fig:chieff} shows the inferred distribution of $\chi_{\rm eff}$ for two choices of the maximum value of $a$ ($1$ and $0.5$).

\begin{figure}
\begin{center}
\includegraphics[width=6.0cm,angle=-90]{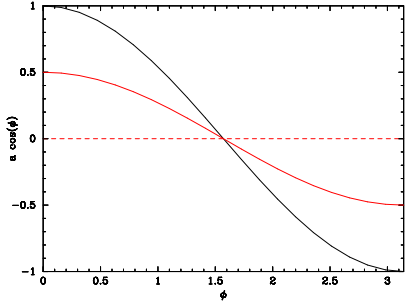}
\end{center}
\caption[acostheta vs theta]{Allowed envelope for $\chi_{\rm eff}$ for binaries dominated by an individual BH embedded in an AGN disk drawn from a population of uniform flat spin $a=[-1,+1]$ where the spin vectors  are oriented at angle $\phi$ compared to the disk gas. The first generation of mergers in this channel will have $\chi_{\rm eff}$ drawn from this distribution. The angle $\phi$ is drawn from a uniform distribution $\phi\in[0,\pi/2]$rad for $a>0$ and $\phi\in[\pi/2,\pi]$rad for $a<0$. Black solid line corresponds to the envelope of the distribution of $\chi_{\rm eff}$ as a function of $\phi$. Red solid line corresponds to median $\chi_{\rm eff}$ as a function of $\phi$ but would also correspond to the envelope of $\chi_{\rm eff}$ for a low natal spin magnitude distribution concentrated between $a\in[-0.5,+0.5]$.
\label{fig:acos}}
\end{figure}

To better understand why even BH spin distributions which individually weakly favor alignment produce $\chi_{\rm eff}$ preferentially near zero, we consider a simplifying limit, where the more massive  BH (labelled ``1'' below) dominates $\chi_{\rm eff}$, as usually holds for our power-law mass distributions.
Figure ~\ref{fig:acos} shows the resulting allowed range of $\chi_{\rm eff}\simeq |a_1|\cos(\phi_1)$ for our initial population of BHs as a function of $\phi_1$, bounded by the black and dashed red lines.
%The allowed values of $\chi_{\rm eff}$ for a given angle $\phi_1$ are bound by the solid black curve and the dashed red line in Fig.~\ref{fig:acos}. 
For example a black hole in the AGN disk with initial angle $\phi_1 \sim 0$ would produce, if it dominates a binary, a $\chi_{\rm eff}$ randomly drawn from a uniform distribution between $\chi_{\rm eff} \sim [0,1]$. By contrast, a black hole with initial $\phi \sim 1.0(2.0)$rad  would have $\chi_{\rm eff}$ randomly drawn from a uniform distribution between $\chi_{\rm eff} \sim [0,0.5]([-0.5,0]])$. The solid red line in Fig.~\ref{fig:acos} indicates median $\chi_{\rm eff}$ for the initial black hole population as a function of $\phi$.
%
%\dwy{In this section, I added absolute value signs to all of the $\chi_{\mathrm{eff}}$'s, as I think that's what's meant to be implied.}
From Fig.~\ref{fig:acos} any BHs with  $\phi$ close to $\pi/2$ must produce small $|\chi_{\rm eff}$.
Conversely, if $\phi$ is significantly different than $\pi/2$, roughly half of BHs will have spins and thus $|\chi_{\rm eff}|$ less than 1/2.   
%$1$ and $2$ radians must have $|\chi_{\mathrm{eff}}|<1/2$, and these are  $\sim 1/3$ of our entire initial population.  For BH with misalignment outside this range, about half will be have spins below $1/2$ and thus  $|\chi_{\mathrm{eff}}|$ less than the median value.   This $\sim 2/3$ of BHs drawn from initial distributions that are uniform in both spin and $\phi$ must have $|\chi_{\mathrm{eff}}|<1/2$. 
Thus, we expect that among the first generation of mergers in this channel, $|\chi_{\mathrm{eff}}|$ must be biased towards low values.
%If the spin distribution were itself biased towards low-spin, say due to an initial natal low spin, then the fraction of the first generation of mergers with $|\chi_{\mathrm{eff}}|<1/2$ could be significantly higher than $\sim 2/3$.\\

\subsection{How do we get low $\chi_{\rm eff}$ mergers in this channel?}
\label{sec:how}
In AGN disks, we expect multiple generations of mergers \citep{McK12,McK14}. In the first generation of mergers in this model, the spin magnitudes and orientations are random, and so from Fig.~\ref{fig:acos}, we are drawing from a population that will be biased towards low values of $\chi_{\rm eff}$. Once BHs encounter each other within their mutual Hill sphere, the binaries have $\mathcal{O}($50:50$)$ odds that their orbital angular momentum about its center of mass is prograde or retrograde. Therefore around $1/4$ of first generation mergers will have both black hole spins anti-aligned with the orbital angular momentum, potentially creating a genuinely low-spin black hole (see \S\ref{sec:spins} below). The net spin that results can be low magnitude (see eqn.~(\ref{eq:amerger}) below for dependencies) but will also tend to be aligned or anti-aligned with the disk gas. Additionally, $\sim 1/2$ of first generation mergers will consist of BHs with opposite sign spins, also potentially yielding low spin magnitude BHs. 

Second generation mergers will mainly consist of mergers between 1st and 2nd generation BHs.  The 2nd generation of BH will be  aligned or anti-aligned with the disk gas, but the 1st generation  remains randomly aligned.  Thus, the 1g-2g mergers will have a higher fraction of larger $\chi_{\rm eff}$ mergers, but will still favor low $\chi_{\rm eff}$.  As the generations evolve, collisions between 2nd or 3rd generations will tend to consist of fully aligned or anti-aligned mergers. Around half of the later generation of 2g-2g or higher mergers will consist of anti-aligned spin BH, and therefore a low $\chi_{\rm eff}$. An additional $\sim 1/4$ of later generation mergers will have spins anti-aligned with binary angular momentum. Thus, low $\chi_{\rm eff}$ mergers should still occur, but the distribution should broaden with increasing generation.

\section{Assumptions behind the simulations}
We ran a series of simulations of embedded black hole migrators in 1-d AGN disk models. In this section we discuss the assumptions behind the different components of our simulations, divided according to useful sub-sections.

\subsection{The AGN disk and the BHs in the disk}
We started with two different models for the gas disks, given by the \citet{SG03} and \citet{Thompson05} disk models. The disk models are calculated for a central supermassive black of mass $M=10^{8}M_{\odot}$ \citep{SG03} and $\sim 10^{9}M_{\odot}$ \citep{Thompson05} respectively. We assumed the disk was populated by BHs drawn from an initial mass function $M^{-\gamma}$, with lower and upper bounds given by $[5,50]M_{\odot}$.  We note the process of grind-down hardens the power law index $\gamma$ relative to the physical IMF appropriate to the AGN disk environment \citep{Yang19}. We assumed that the black hole spherical component in the galactic nucleus consisted of $\sim 2 \times 10^{4}$ BHs \citep{Generozov18}, and that a fraction $N_{\rm SBH}$ between $\sim 0.5\%-5\%$ of this component lived in the disk, corresponding to average aspect ratios for the disk models in \citep{SG03} and \citep{Thompson05} respectively.  We assumed that the BHs were all settled in the equatorial plane of the disk, so that any binaries that form would have orbital angular momentum either aligned or anti-aligned with the disk gas orbital angular momentum ($\vec{L}_{\rm bin} = \pm \vec{L}_{\rm disk}$). The BHs are assumed to lie distributed along the 1-d radial density profile of each disk model.

\subsection{Black hole orbits and migration}
We assumed that half the initial black hole population in the disk ($N_{\mathrm{SBH}}$) lay on prograde circular Keplerian orbits and half the initial population lay on retrograde circular orbits. We also assumed that a component of the spherical population of BHs would end up with orbits ground-down into the AGN disk during the run, adding an additional $N_{\mathrm{grind}}$ population, all with prograde orbits but with random spin magnitudes and orientations. We assumed that gas torques cause embedded black hole orbits to change over time as they undergo migration (a change of semi-major axis) within the disk. The so-called Type I migration time assuming a \citet{SG03} model disk is given by \citet{McK17} as
\begin{eqnarray}
\tau_{\mathrm{mig}} &\sim& 0.03 \mathrm{Myr} \left
(\frac{N}{3}\right)^{-1}\left(\frac{R}{10^{3}r_{\mathrm{g}}}\right)^{1/2}\left(\frac{M}{5M_{\odot}}\right)^{-1}\left(\frac{h/R}{0.01}\right)^{2}\nonumber
\\
&& \left(\frac{\Sigma}{10^{7}{
\mathrm{kg} \, \mathrm{m}^{-2}}}\right)^{-1}\left(\frac{M}{10^{8}M_{\odot}}\right)^{3/2}
\label{eq:mig}
\end{eqnarray}
and this is the time for a massive migrator to migrate onto the SMBH through the disk. The equivalent Type I migration time for a typically thinner, but less dense \citet{Thompson05} model disk can be parameterized as 
\begin{eqnarray}
\tau_{\mathrm{mig}} &\sim& 0.3 \mathrm{Myr} \left
(\frac{N}{3}\right)^{-1}\left(\frac{R}{10^{3}r_{\mathrm{g}}}\right)^{1/2}\left(\frac{M}{5M_{\odot}}\right)^{-1}\left(\frac{h/R}{10^{-3}}\right)^{2}\nonumber
\\
&& \left(\frac{\Sigma}{10^{4}{\mathrm{kg} \, \mathrm{m}^{-2}}}\right)^{-1}\left(\frac{M}{10^{9}M_{\odot}}\right)^{3/2}
\label{eq:mig_tqm}
\end{eqnarray}
or approximately $\times 10$ slower migration than \citep{SG03} at the thinnest part of the \citet{Thompson05} disk model. However, as we have shown in previous work, there can be regions of AGN disks where the net torque on a migrator is zero, leading to the formation of a migration trap within the disk, where masses can encounter each other and merge quickly \citep{Bellovary16,Secunda18}. Here we followed \citep{Bellovary16} and assumed that a migration trap lay at $\sim700r_{\mathrm{g}}$ in the \citep{SG03} model disk and at $\sim 500r_{\mathrm{g}}$ in the \citep{Thompson05} model disk \footnote{Note our migration trap locations are a factor of two different from \citep{Bellovary16} since in that paper the migration traps should have been written in units of the Schwarzschild radius $r_{\mathrm{Sch}}=2r_{\mathrm{g}}$}. All BHs on prograde orbits are assumed to migrate inwards at a rate given by eqn.~(\ref{eq:mig}) from orbits lying outside the migration trap. BHs interior to the migration trap (few in number since this is a small fraction of the area of the disk) are assumed to migrate outwards through the disk to the migration trap. Retrograde BHs are assumed to not migrate since the spiral density perturbations generated by retrograde orbiters are $\sim 1\%$ that of the prograde migrators \citep{McK14}, and so we assume migration torques on retrograde orbiters are negligible for the few Myrs an AGN disk might persist.

\subsection{Black hole spins}
\label{sec:spins}
 The BHs were given initial dimensionless spin parameter magnitudes ($a$) drawn from a flat, uniform distribution between $a=[-1,+1]$. The BHs have an angle $\phi$ between their spin vector ($\vec{a}$) and the angular momentum vector of the disk ($\vec{L}_{\rm disk}$). For BHs with prograde spin ($a>0$), the initial value of $\phi$ was drawn from a uniform distribution $\phi=[0,90^{\circ}]$ where $\phi=0^{\circ}$ corresponds to perfect alignment between the prograde spin and $\vec{L}_{\rm disk}$. For retrograde spin BHs ($a<0$), the initial value of $\phi$ was drawn from a uniform distribution $\phi=[90.01^{\circ},180^{\circ}]$ where $\phi=180^{\circ}$ corresponds to perfect anti-alignment with $\vec{L}_{\rm disk}$. 
 By assuming a uniform distribution in $\phi$ rather than $\rm{cos}(\phi)$ we are assuming that the addition of the AGN disk to the nuclear cluster (which dominates the masses of individual embedded objects) has had a minor effect on the initial distribution of spins. Over time, the spins of the initial black hole population are allowed to evolve as gas accretion reduces $|a|$ for the $a<0$ population and increases $|a|$ for the $a>0$ population. Accreting gas mass will also tend to torque the spin into alignment with the disk gas angular momentum, effectively decreasing $\phi$ towards zero (see next section). \\ 

\subsection{Black hole accretion}
\label{sec:acc}
BHs on prograde orbits in the disk were assumed to accrete at a fraction $f_{\rm Edd}$ of the Eddington rate. For most simulations we used $f_{\rm Edd} \sim 1$. As gas accretion occurs we also expect spin alignment with the disk in the limit of $1\%-10\%$ mass growth from the disk \citep{Bogdanovic07}. The mass-doubling time is $\sim 40$Myr at the Eddington accretion rate, so several Myrs of gas accretion should be required in order to align all the BHs spins with the AGN disk. Spin up or down via accretion takes a relatively long time compared to torquing of angle $\phi$, and if AGN disks are short lived, gas accretion will have little impact on the spin magnitude. More important in the context of an AGN disk, it turns out, is the net spin of a merged black hole (see \S\ref{sec:mergers} below). We expect most mergers to occur early on in the disk's lifetime. Early in the AGN lifetime, gas damping is circularizing BH orbits and the number of collision targets is a maximum \citep{McK12,McK14}. We assumed that spinning BHs which were not aligned with the angular momentum of the gas disk (i.e. $\rm{cos}\phi >0$) would experience a torque due to accreting gas, thus driving the black hole spin towards alignment with the gas disk. Using the maximum estimate from \citet{Bogdanovic07} we assumed that $\sim 1\%$ mass accretion at $f_{\rm Edd} \sim 1$ was sufficient to drive a prograde orbiting black hole spin into alignment with the gas disk. We assume the rate of torquing of alignment is uniform over time. While we may be over-estimating the rate of spin alignment by up to an order of magnitude, super-Eddington gas accretion ($f_{\rm Edd} >1$) is also possible for BHs embedded in the disk, so we believe our choice of alignment rate is a reasonable starting point. Throughout, we ignore the effect of black hole accretion on the surrounding gas and so we assume that the migration and gas hardening torques are unaffected by the back-reaction of the radiation and winds. We will build a feedback prescription into future simulations based on the results of numerical experiments.

\subsection{Binary formation and hardening}
We assume that the binary initial fraction is zero. However, by distributing the BHs randomly across the AGN disk, a fraction randomly end up close to each other.
We assume a binary forms when two prograde orbiters lie within their mutual Hill sphere
\begin{equation}
R_{\rm Hill}=R\left(\frac{q_{\rm bin}}{3} \right)^{1/3}
\end{equation}
where $R$ is the semi-major axis of the center of mass of both BHs and $q_{\rm bin}=M_{b}/M_{\rm SMBH}$ is the ratio of the binary mass ($M_{b}=M_{1}+M_{2}$) to the SMBH mass (assumed to be $10^{8}M_{\odot}$). The disk is assumed to extend to a distance of $2 \times 10^{4}r_{\mathrm{g}}$, so two prograde BHs $M_{1}=50M_{\odot}, M_{2}=50M_{\odot}$ both with semi-major axes $R \sim 10^{4}r_{\mathrm{g}}$ form a binary if they lie within $R_{H} \sim 70r_{\mathrm{g}}(R/10^{4}r_{\mathrm{g}})(M_{\rm bin}/100M_{\odot})^{1/3}$ of their center of mass. In principle a binary can form between a retrograde black hole and a prograde black hole if the separation between the two is $\sim q_{\rm bin}R$, but the probability of this occurring is very small.\\

Once a binary forms we assume that about half the binaries randomly form with prograde angular momentum and half form randomly with retrograde angular momentum. In numerical experiments, we have actually found that out of 30 binary formation encounters that we studied, approximately 17/30 were retrograde, 8/30 were prograde and 5/30 were indeterminate (see Secunda et al. 2019, in prep.). So, the ratio of retrograde: prograde binaries may actually be closer to 2:1, but for our simulations we assumed 1:1 throughout. We will explore differences in this ratio in future work.\\
 
The rate of hardening is assumed to follow the \citet{Baruteau11} results where retrograde binaries harden $\sim \times 5$ faster than prograde binaries. In particular, it takes $\sim 200$ orbits around the center of mass for a retrograde binary to halve its semi-major axis. By assuming that this hardening prescription applies throughout the binary lifetime, binaries experience a runaway hardening effect from gas torques and rapidly get driven to the regime of GW-emission domination, whereupon the binary promptly merges. This torque prescription fails after several halvings of the semi-major axis, the binary will be 'hard' compared to the encounter velocity of prograde migrators, so tertiary encounters will become important in hardening the binary to the point where GW emission dominates. Approximately $\sim 10$ such encounters would be required to harden a binary with semi-major axis $(R_{H}/2)$ to GW-dominated merger \citep{Leigh18}. We will investigate stalled binary hardening and the importance of tertiary encounters in future work.\\

\subsection{Mergers}
\label{sec:mergers}
Merging binary black holes produce a remnant, whose final mass and spin reflects the input BH properties and the overall radiative mass and angular momentum losses during merger.   For clarity, we will not employ very accurate but black-box approximations like \cite{Varma19}, instead adopting simplified but transparent approximations.  We shall approximate the merged mass as \citep{Tichy08}
%\dwy{Standard notation for symmetric mass ratio is $\eta$ instead of $\nu$.  We should probably stick with that for clarity.}
\begin{equation}
M_{\rm final}=M_{\rm b} \left(1-0.2\nu -0.208\nu^2(a_{1}+a_{2}) \right)
\end{equation}
 where $\nu=\mu/M_{\rm b}$, the symmetric mass ratio, or $\nu=q_{b}/((1+q_{b})^2)$ where $q_{b}=M_{2}/M_{1}$ is the binary mass ratio. We  assume that the merged spin magnitude is given by \citep{Tichy08} 
\begin{eqnarray}
a_{\rm merger} &\approx& 0.686 \left( 5.04 \nu -4.16 \nu^{2}\right) \nonumber
\\
&+& 0.4\left(\frac{a_{1}}{(0.632+1/q)^{2}} + \frac{a_{2}}{(0.632 + q)^{2}}\right)
\label{eq:amerger}
\end{eqnarray}
In this expression, when calculating the magnitude of the postmerger spin, we assume $\phi_i$ are small. 
 We assume that black hole binaries always form with orbital angular momentum oriented parallel or anti-parallel to the angular momentum of the gas in the disk  so $L_{\pm}=+1(-1)$ is aligned (anti-aligned) with the angular momentum of the gas. Thus, $\phi_{\rm merger}$ is always fully aligned with the disk ($0$rad for $L=+1$) or anti-aligned with the disk ($\pi$rad for $L=-1$). 

When gravitational radiation starts to dominate the BBH inspiral, the BH is still in a sufficiently wide orbit that its orbital angular momentum dominates over the component spin angular momenta.  Moreover, to a good first approximation, the total angular momentum direction is conserved during binary inspiral, and the remnant BH's spin direction points along this same axis.  We therefore assume the remnant BH has perfect alignment with its initial orbital angular momentum direction.  For comparable-mass binaries, the orbital angular momentum at merger dominates, so the remnant angular momentum direction will point in the same direction as the orbit, independent of BH spins.  Conversely, for highly asymmetric binaries, the orbital angular momentum at merger is much smaller than the binary's spin angular momentum, and the spin angular momentum direction is unchanged.  This effect is  captured in Eq. (\ref{eq:amerger}), above.
%\ros{Dan, check against reliable expressions;  scales as mass squared, not linearly, and may not work well if the angles are asymmetric (e.g., one zero and one pi). Also addition is 2d vectors, not as angles}
%After a merger we assume the angle $\phi$ between $\vec{a}_{\mathrm{merger}}$ and $\vec{L}_{\mathrm{disk}}$ is given by the mass- and spin-weighted quantity
%\begin{equation}
%\phi_{\mathrm{merger}} =\phi_{1}|a_{1}|\frac{M_{1}}{M_{b}} + \phi_{2}|a_{2}|\frac{M_{2}}{M_{b}}.
%\end{equation}

The merging BH experiences a recoil kick which only very slightly perturbs its orbit, given the BH's large Keplerian orbital velocity around the central SMBH.  Because our BH spins are largely aligned or antialigned with the disk, the kick magnitude never reaches the most extreme options possible, nor does it misalign the remnant's orbit from the disk.  For the SMBH and BH of interest, recoil kicks would perturb orbital eccentricity by a small amount (i.e., ${\rm v}_{\rm recoil}/{\rm v}_{\rm Kep} \ll 1$). Since orbital eccentricity damping is very fast in AGN disks \citep{McK12} we expect recoil to not significantly impact the merger or evolutionary history of BHs in the disk (barring exceptionally high numbers of BHs in the disk, so BHs are often separated by only a few Hill radii).   We therefore ignore BH recoil kicks in all subsequent discussion.

\subsection{Grind-down population}
When BHs lie on orbits inclined to the disk they are expected to experience a head-wind and drag force as they pass through the disk. The net result is to extract energy from the black hole orbit and grind its inclination down into the plane of the disk over time \citep[e.g.][]{Syer91,Subr05,McK14}. Grind down of orbits is most efficient at small radii, so we assumed that some of the spherical component of BHs is ground down into the disk at radii $<5000r_{\mathrm{g}}$ over time.  We typically assumed $N_{\mathrm{grind}} \sim 100$/Myr, corresponding to $\mathcal{O}(10\%)$ of a typical initial disk population.

The grind-down rate contributes directly to the rate of first-generation mergers drawn from this population, as well as providing the seeds for all future generations of mergers.

\subsection{Caveats}
One of the biggest caveats for our simulations is that we ignore all tertiary encounters. Tertiary encounters can both help and hinder binary formation. For example, if binaries can only be hardened by gas torques to a modest separation and stall thereafter, the resulting population of hard binaries can become hardened to merger via $\mathcal{O}(10)$ tertiary encounters \citep{Leigh18}. This is because the relative velocity of a prograde encounter is less than the binary velocity around its own center of mass \citep{SigPhin}. Conversely, encounters between a binary and either the spherical component of BHs plunging through the disk or a black hole on a retrograde orbit can ionize wide binaries, since the relative velocity of the encounter is large \citep{Leigh18}.\\

We ignore scattering between retrograde and prograde BHs upon close encounters, although we have started numerical studies of this effect (Secunda et al. 2020, in prep.). We also assume that the retrograde BHs have zero orbital eccentricity. This is unlikely. BHs on prograde orbits have initial eccentricities efficiently damped in a short time ($<0.1$Myr) \citep{McK12}. However, BHs on retrograde orbits are unlikely to have their initial eccentricities damped by much, if at all. As a result, the rate of encounter between retrograde and prograde orbiters could be higher than presented here, with obvious consequences for the rate of three-body encounters and binary ionization as well as scattering events. A future version of this simulation will build in scattering events based on numerical studies of prograde and retrograde encounters (Secunda et al. 2019, in prep).\\

We also ignore complicated multiple encounters within a Hill sphere. Particularly early on, several BHs may lie within a mutual Hill sphere. For now our primary assumption is that the BH pair separated by the smallest 1-d radial distance form a binary and ignore the complexity of interactions with other nearby BHs, although we also later test an alternative scenario where the most massive BH within a mutual Hill sphere form a binary. Furthermore, since our disk model is 1-d, we ignore 2-d and 3-d complications like resonances or slightly inclined orbits. We also ignore precession effects. These are problems we will tackle in future extensions of this work.\\

Finally, we ignore objects other than BHs. There should be a large population of other stellar remnants and stars embedded in the disk \citep{McK12} also subject to similar gas torques. This will lead to binary formation between neutron stars, white dwarfs, stars and BHs. If gas torques are efficient at merging such binaries, electromagnetic signatures may result \citep{McK17}. However, if gas torques lead to stalled hard binaries, BHs in binaries with non-BHs may end up swapping partners in tertiary encounters. Such encounters will be the subject of future numerical work.

\section{Simulation results}
In this section we show results from individual runs to illustrate some of the possible variations within our model. We shall then outline the Monte-Carlo results of a large number of runs, characterizing the parameter distributions. In Table~\ref{tab:runs} we list a range of individual runs to illustrate our choice of particular input parameters.
In Table~\ref{tab:runs_results} we list key results from each of the individual runs.
\begin{table*}
 \begin{minipage}{120mm}
   \caption{Sample runs to illustrate the variations that arise with changing input parameters.Column 1 is the run name. Column 2 is the initial number of BH embedded in the disk. Column 3 is the number of BH ground down by the disk per Myr. Column 4 is the IMF index $M^{-\gamma}$. Columns 5(6) are the lower (upper) bounds to the IMF. Column 7 is the AGN lifetime in Myrs. Column 8 is the spin distribution (u=uniform, flat between $a=[-1,+1]$). Column 9 is the position of the migration trap if present ($700r_{\mathrm{g}}$ in the \citep{SG03} disk model and $250r_{\mathrm{g}}$ in the \citep{Thompson05} disk model. Column 10 shows the choice of disk model (SG=\citep{SG03}; TQM=\citep{Thompson05}).\label{tab:runs}. Column 11 shows the ratio of binary hardening timescale for retrograde ($t_{-}$) versus prograde ($t_{+}$) binaries, with $t_{-}/t_{+}=5$ the result from \citep{Baruteau11}.}
\begin{tabular}{@{}lrrrrrrrrrr@{}} 
\hline
Run & $N_{\mathrm{BH}}$ & $N_{\mathrm{gr}}$ & $\gamma$ & $M_{\mathrm{Lower}}$ & $M_{\mathrm{Upper}}$ & $\tau_{\mathrm{AGN}}$ & $a$ & trap & disk & $t_{-}/t_{+}$\\
       &               &    (/Myr) &  & ($M_{\odot}$) & ($M_{\odot}$) & (Myr) & & ($r_{\mathrm{g}}$)& &\\
\hline
R1 & 869 & $10^{2}$ & 1 & 5 & 50 & 1 & u & $700r_{\mathrm{g}}$ & SG & 5\\
R2 & 869 & $10^{2}$ & 1 & 5 & 50 & 1 & u & $700r_{\mathrm{g}}$ & SG & 1\\
R3 & 100 & $10^{2}$ & 1 & 5 & 50 & 1 & u & $700r_{\mathrm{g}}$ & SG & 5\\
R4 & 851 & $10^{2}$ & 2 & 5 & 50 & 1 & u & $700r_{\mathrm{g}}$ & SG & 5\\
R5 & 851 & $10^{2}$ & 2 & 5 & 50 & 5 & u & $700r_{\mathrm{g}}$ & SG & 5\\
R6 & 851 & $10^{2}$ & 2 & 5 & 15 & 1 & u & $700r_{\mathrm{g}}$ & SG & 5\\
R7 & 851 & $10^{2}$ & 2 & 5 & 50 & 1 & (1-a) & $700r_{\mathrm{g}}$ & SG & 5\\
R8 & 851 & $0$ & 2 & 5 & 50 & 1 & u & $700r_{\mathrm{g}}$ & SG & 5\\
R9 & 851 & $0$ & 2 & 5 & 50 & 5 & u & $700r_{\mathrm{g}}$ & SG & 5\\
R10 & 851 & $0$ & 2 & 5 & 50 & 5 & u & $700r_{\mathrm{g}}$ & SG & 1\\
R11 & 851 & $10^{2}$ & 2 & 5 & 50 & 1 & u & none & SG & 5\\
R12 & 851 & $10^{2}$ & 2 & 5 & 50 & 1 & u & $500r_{\mathrm{g}}$ & TQM & 5\\
\hline
\end{tabular}
\end{minipage}
\end{table*}

\subsection{Initial results from individual disk runs}
Table~\ref{tab:runs} lists twelve fiducial runs (R1-R12) that illustrate a range of possible scenarios for merging stellar mass black holes in AGN disks. In Table~\ref{tab:runs_results} we list some of the key results from the fiducial runs. We divide the results in Table~\ref{tab:runs_results} into black holes merging in the bulk of the disk and the black holes merging at a migration trap (if present). For context, the average mass ratio ($\overline{q}$) and $\overline{\chi}_{\rm eff}$ of the first ten black hole mergers detected with LIGO are $\overline{q} \sim 0.7$ and  $\overline{\chi}_{\rm eff} \sim 0.06$ \citep{LIGO18}. 

\begin{figure}
\begin{center}
\includegraphics[width=6.0cm,angle=-90]{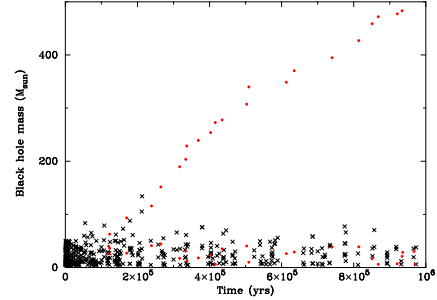}
\end{center}
\caption[Mass of mergers vs time]{Masses of BHs involved in mergers over time from R1. Black crosses are BHs involved in mergers not in the migration trap. Red filled circles are BHs involved in mergers in the migration trap. Of 252 mergers total in the run, 227 mergers occur away from the migration trap in the bulk of the disk and 25 mergers occur at the migration trap. More than half the mergers in this run (121) occur in the first $\sim 0.1$Myr, with 99 in the first $0.05$Myr. Mergers at the migration trap only begin after 0.12Myr.
\label{fig:mt}}
\end{figure}

We illustrate our results by discussing outcomes of representative runs. In R1 from Table~\ref{tab:runs} we had $N_{\mathrm{SBH}}=869$ BHs initially, drawn from an IMF with $\gamma=-1$, and there are a further $N_{grind}=100$ added during the 1Myr disk lifetime. Fig.~\ref{fig:mt} shows the black hole masses involved in mergers over time. Black crosses correspond to mergers occurring in the disk away from the migration trap. Red filled-in circles correspond to mergers at the migration trap and reveal the rapid and steady growth of the IMBH at the migration trap. A majority of mergers in the bulk ($>50\%$) occur early on ($<0.1$Myr) as BHs that happen to lie close to each other in the disk merge quickly. We estimate that our initial random radial distribution was equivalent to an effective initial binary fraction of $\sim 0.15$ in R1. Following this burst of mergers, the number of possible targets for subsequent merger then drops and BHs have to migrate in order to find new targets for merger.

\begin{figure}
\begin{center}
\includegraphics[width=6.0cm,angle=-90]{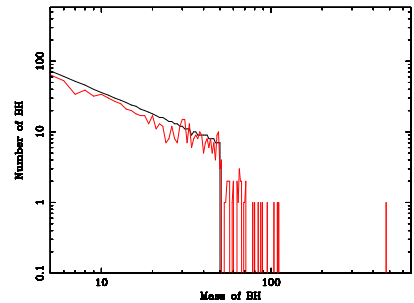}
\end{center}
\caption[Mass spectrum]{Evolution of mass spectrum of BHs in the disk as a result of mergers and gas accretion for R1. Black solid line corresponds to input initial mass function (IMF). Red solid line corresponds to the mass spectrum of BHs in the disk after 0.95Myr. Ground-down BHs added to the disk over time are drawn from a distribution identical to that for the IMF. 
\label{fig:mass_spec}}
\end{figure}

The evolution of the mass spectrum for R1 is shown in Fig.~\ref{fig:mass_spec}. The black solid line corresponds to the input initial mass function (IMF) and the red solid line corresponds to the mass function after 1Myr. The IMF has evolved towards a broken powerlaw distribution, with the break lying near the high mass end of the IMF \citep{McK17}, along with a single IMBH.  Note that the low mass end of the distribution is  supported by the addition of ground-down BHs drawn from a distribution identical to the IMF.  

\begin{table*}
 \begin{minipage}{150mm}
   \caption{Initial results from sample runs. Column 1 is the run name. Column 2 is the number of mergers in the bulk of the disk during the run. Column 3 is the number of mergers at the trap (if present). Column 4 is the ratio of bulk mergers involving at least $1$ BH of 2g or higher, or $ngmg/1g1g$ where $n>1,m \geq 1$. Column 4 is the median mass ratio ($\tilde{q}$) per bulk merger with associated standard deviation. Column 5 shows the range of $q$ for bulk mergers during the run. Column 6 is the median $\tilde{\chi}_{\rm eff}$ for mergers in the bulk, with associated standard deviation. Column 7 shows the range of $\chi_{\rm eff}$ for the bulk mergers during the run. Column 8 shows the largest mass BH in the run (mostly at trap). Column 9 shows the time of 1st merger at the trap and Column 10 the mass of the 1st merger at the trap \label{tab:runs_results} 
   %\dwy{I think it'd look cleaner if the columns with two values (e.g., $[q_{\mathrm{min}},q_{\mathrm{max}}]$) were split into two columns.}
   }
\begin{tabular}{@{}lrrrcrcrrrr@{}} 
\hline
Run & $N_{\rm bulk}$ & $N_{\rm trap}$ &$ngmg/1g1g$ &$\tilde{q}$ & [$q_{\rm min},q_{\rm max}$] & $\tilde{\chi}_{\rm eff}$ & [$\chi_{\rm min},\chi_{\rm max}$] & $M_{\rm IMBH}$ &$t_{\rm trap}$ & $M_{\rm trap}$ \\
 & & & & & & & & ($M_{\odot}$) & (Myr) & ($M_{\odot}$)\\
\hline
R1 & 227 & 25 & 0.22 &$0.46\pm 0.24$& [0.08,1.0] & $0.07\pm 0.37$ & [-0.88,0.83] & 552.7 &0.12 & 63.0\\
R2 & 189 & 26 & 0.16 & $0.44\pm 0.25$& [0.06,1.0] & $0.08\pm 0.40$ & [-0.94,0.87] & 745.0 & 0.12 & 73.0 \\
R3 & 16 & 21 &0.19 &$0.47\pm 0.20$ & [0.18,1.0] & $-0.09\pm 0.37$ & [-0.83,0.56] & 486.0 & 0.38 & 46.6 \\
R4 & 172 & 17& 0.15&$0.63\pm 0.27$ & [0.10,1.0] & $0.03\pm 0.33$ & [-0.79,0.88] & 272.2 & 0.24 & 42.1 \\
R5 & 329 & 159 & 0.29 &$0.50\pm0.25$ & [0.07,1.0] & $0.03\pm0.44$ & [-0.96,0.94] & 2409.7 & 0.14 & 30.2 \\
R6 & 175 & 14 & 0.09 &$0.70\pm 0.19$ & [0.26,1.0] & $-0.03\pm 0.34$& [-0.79,0.81] & 126.6 & 0.19 & 12.1 \\
R7 & 178 & 17 & 0.18&$0.57\pm 0.26$ & [0.09,1.0] & $0.01\pm 0.23$ & [-0.59,0.82] & 270.2 & 0.24 &42.7  \\
R8 & 157 & 1 & 0.08 &$0.60 \pm 0.26$ & [0.11,1.0] & $0.07\pm 0.37$ & [-0.86,0.88] & 16.3 & 0.24 & 16.3 \\
R9 & 221 & 27 &0.27 &$0.50 \pm 0.26$ & [0.07,1.0] & $0.00\pm 0.44$ & [-0.94,0.94] & 709.8 & 0.24 & 16.3 \\
R10 & 221 & 20 &0.25 &$0.49 
\pm 0.26$ & [0.05,1.0] & $0.14\pm 0.46$ & [-0.93,0.93] & 503.4 & 0.24 & 24.1 \\
R11 & 181 & 0& 0.18&$0.54 \pm 0.27$ & [0.05,1.0] & $0.00\pm 0.38$ & [-0.89,0.89] & 139.4 & none & none  \\
R12 & 53 & 149& 0.08&$0.64 \pm 0.24$ & [0.14,1.0] & $0.00\pm 0.31$ & [-0.68,0.64] & 1625.5 & 0.06 & 36.2  \\
\hline
\end{tabular}
\end{minipage}
\end{table*}

\begin{figure}
\begin{center}
\includegraphics[width=6.0cm,angle=-90]{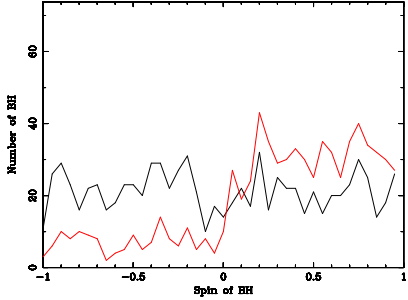}
\end{center}
\caption[Spin spectrum]{Evolution of spin spectrum of BHs in the disk as a result of mergers and gas accretion for R1. Black solid line corresponds to input initial spin distribution, drawn from a uniform, flat distribution. Red solid line corresponds to the spin spectrum of BHs in the disk after 0.95Myr. Ground-down BHs added to the disk over time are drawn from a spin distribution identical to the initial spin  distribution. The final distribution has not been corrected for spin flips due to retrograde binary orbital angular momentum. The angle $\phi$ contains that information in our simulations. 
\label{fig:spin_spec}}
\end{figure}

The spins of the BHs also evolve in R1 over time. Fig.~\ref{fig:spin_spec} shows the evolution of the spins of the initial population on prograde orbits after 1Myr. The black solid line in Fig.~\ref{fig:spin_spec} shows the initial distribution of spins drawn from a uniform, flat spin distribution. The red solid line in Fig.~\ref{fig:spin_spec} shows the final state of the spin distribution among this population after 1Myr. The retrograde orbiters are assumed to not accrete from the gas at any significant rate and their rate of interaction with migrating prograde orbiters is very small. Note that the final distribution only gives the magnitude of the spin, it does not take into account the fact that $\phi$ may have flipped sign (i.e. that the spin points in the opposite direction). 

\begin{figure}
\begin{center}
\includegraphics[width=6.0cm,angle=-90]{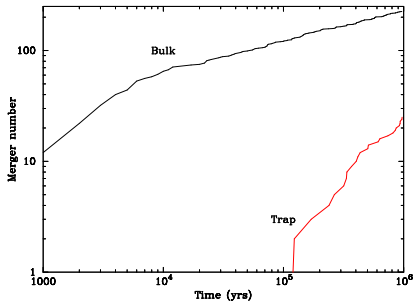}
\end{center}
\caption[Mass spectrum]{The integrated number of black hole mergers as a function of time in R1 for bulk (black) and trap (red). A majority of mergers ($>50\%$) occur early on ($\leq 0.1$Myr). The initial number of mergers grows as $\sim t^{1}$ at $<0.01$Myr and then grows as $\sim t^{1/4}$ at $>0.01$Myr. The population of ground-down orbits provides a support for the merger rate at later times. 
\label{fig:mergers_t} }
\end{figure}

Fig.~\ref{fig:mergers_t} shows the number of black hole mergers as a function of time in R1. As can be seen from Fig.~\ref{fig:mt} most mergers occur early on ($<0.1$Myr). We assumed an initial binary fraction of zero, but the random distribution of BHs in the disk corresponds to an effective initial binary fraction of $f_{\rm bin} \sim 0.15$. These BHs merge quickly in our prescription but then must migrate within the disk to find more partners. Thus, we find a cascade of mergers (growing $\propto t^{1}$) in the first $\sim 0.01$Myr and then growing more like $\propto t^{1/4}$ from 0.01-1Myr. This may suggest that AGN disks are most efficient at black hole mergers early on in their lifetimes. A somewhat counter-intuitive point then emerges: if AGN disks are short-lived, they may end up increasing the rate of black hole mergers detectable with LIGO. This is because shorter-lived AGN disks imply a large rate of AGN episodes per galactic nucleus (for a constant fraction of active nuclei per volume). The AGN 'grinder' then gets multiple opportunities to accelerate mergers of BHs within the overdense central parsec.

\begin{figure}
\begin{center}
\includegraphics[width=6.0cm,angle=-90]{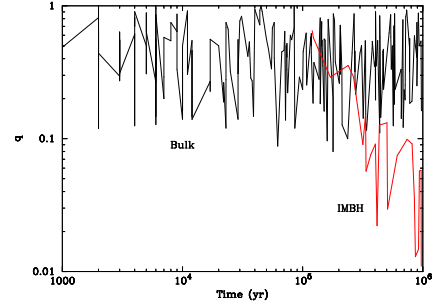}
\end{center}
\caption[Mass ratio]{The mass ratios of mergers as a function of time in R1. In black are the mass ratios of mergers in the bulk of the disk and in red are the mass ratios of mergers at the migration trap. 
\label{fig:r1_q_t} }
\end{figure}

Fig.~\ref{fig:r1_q_t} shows the mass ratio for mergers in R1 as a function of time. In black are mergers away from the merger trap and in red are mergers at the migration trap. The mass ratio of mergers in the bulk population spans $q \sim[0.1,1]$.

\begin{figure}
\begin{center}
\includegraphics[width=6.0cm,angle=-90]{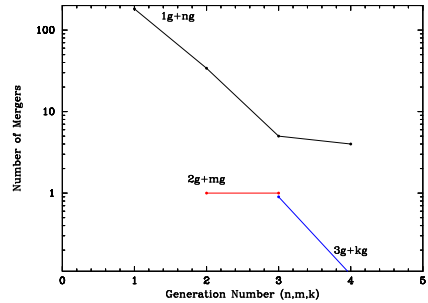}
\end{center}
\caption[Merger generation vs number]{Generations of BHs involved in bulk mergers from R1. Black curve is the 1g$+$ng mergers, where $n \geq 1$. Red curve is 2g$+$mg mergers where $m \geq 2$. Blue curve is 3g$+$kg mergers where $k \geq 3$. For example, there are $38$ 1g$+$2g mergers in this instance of R1. The majority of mergers are by far 1g$+$1g.
\label{fig:gens}}
\end{figure}

Fig.~\ref{fig:gens} shows the generations of BH involved in mergers in the bulk of R1. 1g$+$1g mergers dominate ($78\%$), with 1g$+$2g mergers making up most of the rest ($17\%$) of the mergers in R1.

\begin{figure}
\begin{center}
\includegraphics[width=6.0cm,angle=-90]{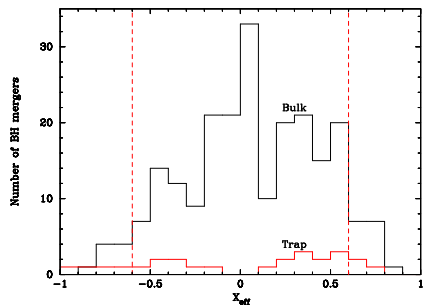}
\end{center}
\caption[Mass spectrum]{The distribution of $\chi_{\rm eff}$ for the black hole mergers in R1. Black solid curve is the $\chi_{\rm eff}$ distribution for mergers in the bulk of the disk. Red solid curve is the distribution for mergers at the trap, and hints at bi-modality. Vertical red dashed lines correspond to the bounds of 90$\%$ of the mergers. The bulk $\chi_{\rm eff}$ distribution appears to be approximately centered around $\chi_{\rm eff} \sim 0$. 
\label{fig:chi}}
\end{figure}

Finally, Fig.~\ref{fig:chi} shows the distribution of $\chi_{\rm eff}$ for all the mergers in R1. The $\chi_{\rm eff}$ distribution given by the black solid curve corresponds to the distribution for mergers in the bulk of the disk. The red solid line corresponds to the distribution of $\chi_{\rm eff}$ for mergers at the trap. Both merger distributions  appear centered approximately around $\chi_{\rm eff} \sim 0$, with the possibility of a bimodal distribution in $\chi_{\rm eff}$ in the trap distribution. Red vertical dashed lines confine $\approx 90\%$ of the bulk distribution. Our initial results confirm our expectations from \S\ref{sec:low_chi} that the $\chi_{\rm eff}$ distribution should be biased towards $\chi_{\rm eff} \sim 0$ at least in the bulk distribution, although we require a larger scale simulation (below) to confirm this statistically.

\subsection{The effect of changing input parameters}
From Table~\ref{tab:runs_results} we varied a number of our input parameters to illustrate the impact of changes in individual parameters. These results are merely to help guide the reader in what to expect from the full Monte Carlo runs. Here we note some of the gross features of the results, but refrain from discussion of the distributions until we have much larger samples.\\

In R2 we changed the ratio of gas hardening timescales between retro- and pro-grade binaries from $t_{-}/t_{+}=5$ \citep{Baruteau11} in R1  to $t_{-}/t_{+}=1$. As a result, we remove a source of very rapid mergers early on and so it is no surprise the number of mergers in the bulk of R2 is lower than in R1. In R3 we reduced the number $N_{BH}$ from R1 and in Table~\ref{tab:runs_results} we can see a much lower number of mergers. The IMBH takes longer to start assembling at the migration trap in R3 compared to R1. However, in the end, R3 yields a similar mass IMBH to that in R1.\\

In R4 we changed the IMF index to $\gamma=2$. The changed powerlaw normalization led to a slightly smaller $N_{BH}=851$ compared to R1. After 1Myr we find a similar number of mergers, but a smaller IMBH and higher $\tilde{q}$ than in R1 or R3. In R5, we extended the R4 run to 5Myr, resulting in substantially more mergers in the bulk of the disk, but an order of magnitude increase in number of mergers at the trap. As a result, R5 yields the largest IMBH of all the runs. Note that the IMBH mass growth from $\sim 30M_{\odot}$ to $\sim 2000M_{\odot}$ in 5Myr corresponds to $>6$ mass doublings in $\sim 1/8$ Eddington mass-doubling times. Thus, the IMBH grows at a remarkable $\times 40-50$ Eddington via collisions, as we expect \citep{McK12}. Also in R5, the fraction of hierarchical mergers (ngmg/1g1g, where $n>1$,$m\geq1$) is higher in the bulk than in runs R1-R4.\\

In R6, we changed the IMF to match the masses of BHs observed in our own Galaxy ($[5,15]M_{\odot}$) which yields a low-mass IMBH after 1Myr and higher $\tilde{q}$. The low fraction of hierarchical mergers could correspond to the generally lower rates of migration.  In R7 we changed the initial spin magnitude distribution to $a(1-|a|)$, where $a$ is drawn from $a=[-1,+1]$. This narrower spin distribution could physically correspond to a low natal spin population of BHs in the IMF. This results in a much narrower $\chi_{\rm eff}$ distribution. In R8 and R9 we investigated the effect of grind-down of orbits into the disk. By removing grind-down, the black hole mass spectrum is no longer supported at the low-mass end with time. Also in R8 after 1 Myr, from Table~\ref{tab:runs_results}, there is no IMBH. This suggests that orbital grind-down at small radii is a key driver of early IMBH formation at a migration trap. In R9 and R10, we extend the disk lifetime from R8 to 5Myr, keeping no grind-down,  and now we grow an IMBH at the trap. Thus, if $\tau_{AGN}$ is long enough, a large IMBH mass can build at a migration trap, independent of grind-down efficiency. The fraction of hierarchical mergers in the bulk increases substantially with no grind-down and longer disk lifetime.

In R11 we removed the migration trap. This prevents the build up of a single massive IMBH, with the largest post-merger BH only $\sim 139M_{\odot}$. 

\begin{table*}
 \begin{minipage}{140mm}
   \caption{Large-scale runs to evaluate parameter distributions for mergers in the bulk of the disk. Column 1 lists the run from Table~{\ref{tab:runs}}, which is then run with different numerical seeds $10^{2}$ times. Column 2 lists the median mass ratio ($\tilde{q}$) and standard deviation in the merger population. Column 3 lists the full range of $q$ for each run. Column 4 lists the median $\chi_{\rm eff}$ and standard deviation for the merger population. Column 5 lists the full range of $\chi_{\rm eff}$ for the run. Column 6 lists the median mass and standard deviation of black hole masses involved in mergers. If the standard deviation is greater than the difference between the median mass and the minimum mass in the run, we list the latter instead as the lower limit. Column 7 lists the maximum merged mass and in brackets maximum merged mass when largest mass merger occurs within a Hill sphere. Column 8 lists the total number of mergers for the run. Column 9 lists the percentage of black holes with mass $>50M_{\odot}$ involved in all mergers, with in brackets the percentage when largest mass merger occurs within a Hill sphere. 
    \label{tab:bigruns_disk}}
\begin{tabular}{@{}lcrcrcrrc@{}} 
\hline
Run & $\tilde{q}$ & $[q_{\rm min},q_{\rm max}]$& $\tilde{\chi}_{\rm eff}$& $[\chi_{\rm min},\chi_{\rm max}]$ & $\tilde{M}_{\rm merger}$ &$M_{\rm max}$ &  $N_{\rm mergers}$ &$N_{>50M_{\odot}}$\\
       &               &     &  & & ($M_{\odot}$) &($M_{\odot}$) & &($\%$)\\
\hline
R1 & $0.48\pm 0.25$ & [0.02,1.0] & $-0.01 \pm 0.33$ & [-0.98,0.98] &$16.3^{+15.0}_{-11.3}$& 258.6(270.5) &17,601 & 1.0(3.0)\\
R2 & $0.48\pm 0.25$ & [0.08,1.0] & $0.00 \pm 0.33$ & [-0.98,0.97] &$17.0^{+13.4}_{-12.0}$& 93.5(236.3) &15,563 & 1.0(3.0)\\
R3 & $0.57\pm 0.26$ & [0.10,1.0] & $-0.04\pm 0.34$ & [-0.91,0.84]& $14.0^{+12.8}_{-9.0}$& 92.7(135.2) & 761 & 0.1(3.0)\\
R4 & $0.56 \pm 0.25$ & [0.10,1.0]& $-0.01 \pm 0.32$ & [-0.95,0.98]& $9.0^{+9.1}_{-4.0}$& 87.1(164.9) &14,963 & 0.1(0.7)\\
R5 & $0.56 \pm 0.24$ & [0.10,1.0] &$-0.02 \pm 0.43$  &[-0.98,0.98] & $9.0^{+22.8}_{-4.0}$ & 144.3(243.3) & 29,405 &0.2(1.0)\\
R6 & $0.71 \pm 0.19$ & [0.33,1.0] & $-0.03 \pm 0.31$ & [-0.98,0.95] & $7.0^{+2.9}_{-2.0}$& 33.3(90.7) & 14,298 & 0.0(0.1)\\
R7 & $0.56 \pm 0.25$ & [0.10,1.0] &$-0.01 \pm 0.11$ & [-0.34,0.64] & $9.0 ^{+9.2}_{-4.0}$& 89.8(151.0) & 15,004 & 0.1(0.8)\\
R8 & $0.56 \pm 0.25$ & [0.10,1.0] & $-0.02 \pm 0.33$ & [-0.97,0.94] & $9.0^{+9.3}_{-4.0}$& 91.8(219.4) & 14,001 & 0.2(0.6)\\
R9 & $0.55 \pm 0.24$ & [0.10,1.0] & $0.00 \pm 0.41$ & [-0.98,0.98] & $9.0^{+9.7}_{-4.0}$& 83.8(183.0) & 19,716 & 0.3(0.9)\\
R10 & $0.55 \pm 0.24$&[0.10,1.0] & $0.05 \pm 0.42$  & [-0.98,0.98] & $9.0^{+9.5}_{-4.0}$&92.4(412.0) &18,276 & 0.2(0.9)\\
R11 & $0.56 \pm 0.25$ & [0.10,1.0] & $-0.01 \pm 0.32$ & [-0.98,0.97] & $9.0^{+9.2}_{-4.0}$& 93.8(201.0) & 15,311 & 0.2(0.7)\\
R12 & $0.64 \pm 0.25$ & [0.05,1.0] & $-0.10 \pm 0.32$ & [-0.98,0.94] & $8.1^{+13.2}_{-3.1}$& 97.8(120.8) &3,992 & 0.2(0.5)\\
\hline
\end{tabular}
\end{minipage}
\end{table*}

\begin{table*}
 \begin{minipage}{130mm}
   \caption{Large-scale runs to test parameter distributions for mergers at the migration trap (where it exists). Column 1 lists the run from Table~{\ref{tab:runs}}, which is then run $10^{2}$ times. Column 2 lists the median mass ratio ($\tilde{q}$) from mergers and the associated standard deviation in the merger distribution. If the standard deviation is greater than the lower bound for the run, we use the lower bound for the run instead. Column 3 lists the median $\chi_{\rm eff}$ and standard deviation for the mergers. Column 4 lists the full range of $\chi_{\rm eff}$ for the run. Column 5 lists the median mass of black holes involved in mergers at the trap together with the standard deviation. Column 6 lists the maximum mass in mergers at the migration trap. Column 7 lists the median masses of the incoming BH merging with the mass at the trap, with the standard deviation of that distribution. If the standard deviation is greater than the difference between the median mass and the minimum mass in the run, we list the latter instead as the lower limit. Column 8 lists the total number of mergers at the migration trap for the run. 
    \label{tab:bigruns_trap}}
\begin{tabular}{@{}lccccccr@{}} 
\hline
    Run & $\tilde{q}$ & $\tilde{\chi}_{\rm eff}$& $[\chi_{\rm min},\chi_{\rm max}]$ & $\tilde{M}_{\rm trap}$ &$M_{\rm max}$ & $\tilde{M}_{\rm merge}$ & $N_{\rm mergers}$\\
                    &     &  & & ($M_{\odot}$) &($M_{\odot}$) & ($M_{\odot}$) &\\
\hline
R1 & $0.08^{+0.19}_{-0.07}$ & $0.01 \pm 0.52$ & [-0.96,0.98] &$263.6\pm 157.7$& 1255.8 & $19.2^{+16.0}_{14.2}$&2,333\\
R2 & $0.09^{+0.18}_{-0.08}$ & $-0.06\pm 0.51$ & [-0.96,0.96]& $262.6\pm 141.5$& 677.0 & $19.2^{+14.7}_{-14.2}$& 2,349\\
R3 & $0.15^{+0.21}_{-0.14}$ & $-0.05 \pm 0.48$ & [-0.97,0.96]& $164.5\pm 90.2$& 493.5 & $19.5\pm 13.9$ &1,368\\
R4 & $0.12^{+0.21}_{-0.11}$ & $0.06 \pm 0.52$  &[-0.96,0.96] & $119.1\pm 65.9$ & 330.0 &$11.1^{+10.2}_{-6.1}$ & 1,494\\
R5 & $0.02^{+0.09}_{-0.01}$ & $0.00 \pm 0.59$ & [-0.98,0.98] & $996.3 \pm 625.7$& 2469.1 &$12.1^{+11.8}_{-7.1}$ & 15,015\\
R6 & $0.16^{+0.21}_{-0.15}$ & $0.04 \pm 0.48$ & [-0.95,0.95] & $62.1 \pm 33.8$& 168.6 & $8.0^{+3.2}_{-3.0}$ &1,409\\
R7 & $0.13^{+0.22}_{-0.12}$ & $0.01 \pm 0.38$ & [-0.76,0.79] & $116.3 \pm 71.8$& 418.3 & $11.1^{+10.6}_{-6.1}$ & 1,487\\
R8 & $0.29\pm 0.26$ & $0.09 \pm 0.52$ & [-0.90,0.97] & $58.1 \pm 34.1$& 169.1 & $10.1^{+9.7}_{-5.1}$ & 502\\
R9 & $0.14^{+0.20}_{-0.13}$ & $0.13 \pm 0.62$ & [-0.95,0.98] & $169.7 \pm 138.0$& 920.6 & $19.5^{+17.3}_{-14.5}$ & 1,852\\
R10 &$0.15^{+0.19}_{-0.14}$ &$-0.02 \pm 0.62$  & [-0.94,0.98] &$154.5\pm 125.1$ & 610.0 & $18.7^{+15.4}_{-13.7}$ & 1,819\\
R12 & $0.01^{+0.11}_{-0.01}$ & $0.01 \pm 0.54$ & [-0.98,0.98] & $782.3 \pm 473.9$& 2194.7 & $9.2^{+11.4}_{-4.2}$ & 14,810\\
\hline
\end{tabular}
\end{minipage}
\end{table*}

Finally, in R12, we replaced the disk model of \citet{SG03} with \citet{Thompson05}, which orbits a $10^{9}M_{\odot}$ SMBH and is thinnest around $10^{3}r_{g}$, with generally lower density, leading to a migration time longer on average than in \citet{SG03}. In this disk model, the migration trap lies at $500r_{\mathrm{g}}$, near the thinnest part of the disk where migration is fastest. As a result, the migration trap in a \citet{Thompson05} disk encounters many BH.  Thus, if AGN disks are generally more like the \citet{Thompson05} model rather than the \citet{SG03} model, contain a migration trap and live $\geq$ 1Myr, we expect the formation of many massive IMBH in AGN disks.

One immediate conclusion from the final two columns in Table~\ref{tab:runs_results} is that if disk lifetimes are actually very short ($\sim 0.1-0.2$Myr), there is very little time to build much mass at migration traps. Thus, ongoing upper limits on BH mergers masses from LIGO are very important astrophysically, since this will put strong limits on disk lifetime, disk structure (including the existence of migration traps) and orbital grind-down efficiency.

\subsection{Monte Carlo simulations}
\label{sec:mc}
We ran individual runs R1-R12 for $100$ iterations, each with different initial random numerical seeds, to extract large distributions of BH masses and $\chi_{\rm eff}$. Separately, we also ran versions of runs R1-R12 that incorporated a modified merger condition: where more than two BH lay within a mutual Hill sphere, the two most massive BH were assumed to merge.   Table~\ref{tab:bigruns_disk} shows the results of Monte Carlo simulations of the various individual runs R1-R12 from Table~\ref{tab:runs} for mergers in the bulk of the disk. Table~\ref{tab:bigruns_trap} shows the results of Monte Carlo simulations of the various individual runs for mergers at the migration trap.\\

From Table~\ref{tab:bigruns_disk}, in general, across a wide range of parameterization, the $\chi_{\rm eff}$ distribution for mergers in the bulk of the disk is centered on $\tilde{\chi}_{\rm eff} \approx 0$ with a standard deviation of $\sigma_{\chi} \sim 0.3$. The very small negative bias in runs R1,R3-R8, R10-R12 is due to enhanced production of negative, anti-aligned spin BH by our choice of $t_{-}/t_{+}=5$. This is confirmed in runs R2 and R10, where $t_{-}/t_{+}=1$, and $\tilde{\chi}_{\rm eff}=0.00,0.05$ respectively. The $\chi_{\rm eff}$ distribution is relatively broad, with $\sigma_{\chi} \sim 0.3$ generally. The only exception is in R7, where a narrow distribution of $\chi_{\rm eff}$ with $\sigma_{\chi} \sim 0.1$ corresponds to a narrow, non-uniform, initial spin magnitude distribution. If the distribution of mergers follows a normal distribution, $\sim 68\%$ of mergers have $|\chi_{\rm eff}|<0.3$ for a flat, uniform, distribution of initial spin magnitudes. As LIGO continues to operate, the width of the $\chi_{\rm eff}$ distribution and a net negative bias in $\tilde{\chi}_{\rm eff}$ may allow us to distinguish between a pure dynamics channel and the AGN disk channel.\\

Also from column 6 in Table~\ref{tab:bigruns_disk}, a flatter IMF ($M^{-\gamma}$) with $\gamma \sim 1$ in R1-R3 yields a higher median mass involved in mergers than in R4-R12. Because LIGO is much more sensitive to high-mass binaries, this steeper power-law index produces a detected distribution that is consistent with early LIGO observations; see Figure \ref{fig:PrimaryMass:Empirically}.  Both powerlaw distributions $\gamma=1,2$  are consistent with inferences about the selection-corrected spectrum of BH masses in BH-BH binaries \cite{Abbott18Pop}. From column 7 in Table~\ref{tab:bigruns_disk}, the maximum mass produced in bulk mergers is $O(100-200M_{\odot})$ generally. The higher maximum mass is generally associated with assuming the most massive BH within a mutual Hill sphere will merge. Changing the condition on mergers within the Hill sphere made no significant difference to the parameter distributions in R1-R12 otherwise. This is a reasonable mass upper limit to expect if migration traps do not generally occur in AGN disks. Finally, column 9 in Table~\ref{tab:bigruns_disk} shows the percentage of black holes in all mergers in the bulk with mass $>50M_{\odot}$. If we assume the nearest BH within a mutual Hill sphere merge, then in no run does this percentage rise above $1\%$. If we change the merger condition to require that the two most massive BH within a mutual Hill sphere to merge, then the percentage of mergers involving a merger in the upper mass gap rises by a factor $\sim 3$ for $M^{-1}$ IMF and by a factor $\sim 3-8$ for $M^{-2}$ IMF. Choice of IMF dominates the upper mass-gap percentage of mergers. Nevertheless, it seems most mergers are between the more numerous, lower-mass BH in our distributions. 

From Table~\ref{tab:bigruns_trap}, we can see that the median mass ratio for mergers is usually much smaller at the trap ($\tilde{q} \sim 0.1$) than in the bulk of the disk ($\tilde{q} \sim 0.5-0.7$). We can also see that the median $\chi_{\rm eff}$ is modestly larger at the migration trap than in the bulk of the disk mergers. From \S\ref{sec:low_chi} above, we expect that later generations of mergers should tend towards alignment and anti-alignment, and this seems consistent with the results for mergers at the migration trap. Changing the merger condition within the Hill sphere so that the two most massive BH merge, made no significant difference to results for the trap. Comparing Table~\ref{tab:bigruns_disk} and \ref{tab:bigruns_trap}, we also find the mass of an incoming black hole arriving at the migration trap is generally larger than the median mass of black holes merging in the bulk of the disk. This is because more massive black holes migrate more rapidly than less massive BH (eqn.~\ref{eq:mig}) so we expect that more massive BH should arrive at the migration trap first. The results from Table~\ref{tab:bigruns_trap} seem inconsistent at the $\sim $2$\sigma$ level with the results observed with LIGO so far. Future results from O3 may therefore rule out the presence of migration traps in AGN disks. If this is the case then either AGN disks do not possess steep changes in aspect ratio or density in the inner disk, or AGN disks are generally short-lived ($\leq 10^{5}$yr) unstable configurations, nested within a fuel reservoir or torus.\\

\begin{figure}
\begin{center}
\includegraphics[width=6.0cm,angle=-90]{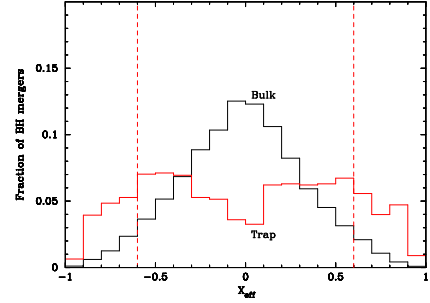}
\end{center}
\caption[Mass spectrum]{Normalized $\chi_{\rm eff}$ distribution for R1 carried out $10^{2}$ times with different numerical seeds. Black solid curve shows the distribution of mergers in the bulk of the disk. Red solid curve shows the distribution of mergers at the trap. Approximately $90\%$ of mergers in the bulk occur between the vertical red dashed lines.  
\label{fig:chi_r1x100}}
\end{figure}

\begin{figure}
\begin{center}
\includegraphics[width=6.0cm,angle=-90]{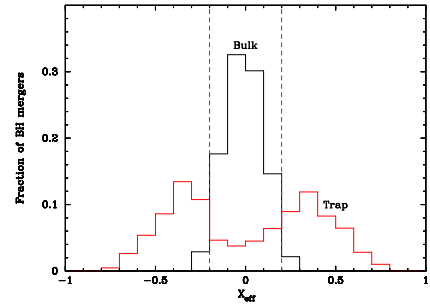}
\end{center}
\caption[Mass spectrum]{As Fig.~\ref{fig:chi_r1x100} except for R7. Approximately $95\%$ of mergers in the bulk of the AGN disk occur between the vertical red dashed lines. 
\label{fig:chi_r6x100}}
\end{figure}

Fig.~\ref{fig:chi_r1x100} shows the total normalized $\chi_{\rm eff}$ distribution from the R1 Monte Carlo run. 90$\%$ of the bulk mergers have $|\chi_{\rm eff}<0.6|$, but the trap distribution seems bimodal, driven by spin flipping of mergers at the trap due to random binary orbital angular momentum. By contrast, R7 is the one run that shows a consistently narrower $\chi_{\rm eff}$ distribution and indeed in Fig.~\ref{fig:chi_r6x100} we see a far narrower distribution of $\chi_{\rm eff}$ among mergers in the bulk of the disk. The trap distribution remains bimodal, but the peaks are also narrower than in Fig.~\ref{fig:chi_r1x100}. In Fig.~\ref{fig:chi_r6x100} $\sim 95\%$ of the bulk mergers occur with $|\chi_{\rm eff}|<0.2$ (between the vertical red dashed lines). As we can see, a narrower spin input distribution (e.g. from a narrow natal spin distribution) clearly leads to a narrow $\chi_{\rm eff}$ distribution in this channel. 

\section{Implications for O3 and beyond}
\label{sec:implications}
From \S\ref{sec:mc} there are several constraints that we hope O3 will provide for this channel. First, a continuing build-up of the $\chi_{\rm eff}$ distribution for binary black hole mergers will allow us to restrict the allowed black hole IMF ($M^{-\gamma},M_{\rm min}, M_{\rm max}$) in galactic nuclei and the allowed initial spin magnitude distribution. Second, a build-up of the merger mass ratio ($q$) distribution and a merging mass upper bound allows us to constrain both the IMF and the possibility of migration traps in AGN disks. This latter constraint, if real, will tell us either: there are no steep gradients in density or aspect ratio in AGN disks, or that AGN disks are typically lower density than in a \citep{SG03} model, or that AGN disks are very short-lived ($\leq 10^{5}$yr) instabilities in a nuclear accretion flow.

We expect that the merger rate from this channel should increase with redshift $(z)$ out to $z\sim 2$ \citep{McK17}, which is the presumed peak redshift for AGN activity, as well as being the peak redshift for (massive) star formation. This expectation is directly testable via LIGO observations. In tandem with LIGO observations, in the electromagnetic band, we can search for IMBH build up in AGN disks as well as EMRIs embedded in AGN disks by searching for Hill sphere shocks \citep{McK19}, periodic wobbles or short-lived ripples in the broad component of the FeK$\alpha$ line \citep{McK13,McK15} or in analogous oscillations in blurred soft X-ray features. Constraints on such effects will allow us to test the distribution of embedded masses in AGN disks as well as the gas torques that we expect to operate there.\\
 
 \section{Gravitational wave phenomenology for the AGN-disk channel}
 
As our understanding of this channel improves and as observations rapidly accumulate, we expect that direct comparisons between evolutionary models and observations will allow us to place sharp constraints on the underlying physics involved.  At present, however, we adopt a phenomenological approach for the time-averaged mass and spin distribution of merging BHs formed in AGN disks.  

\begin{figure}
% python3 plot_distribution_and_model.py --directory ../data/simulations/v3f/ trap  m_1 --x-log-scale --showdist uniform
% python3 plot_distribution_and_model.py --directory ../data/simulations/v3f/ bulk  m_1 --x-log-scale --showdist 1
\includegraphics[width=\columnwidth]{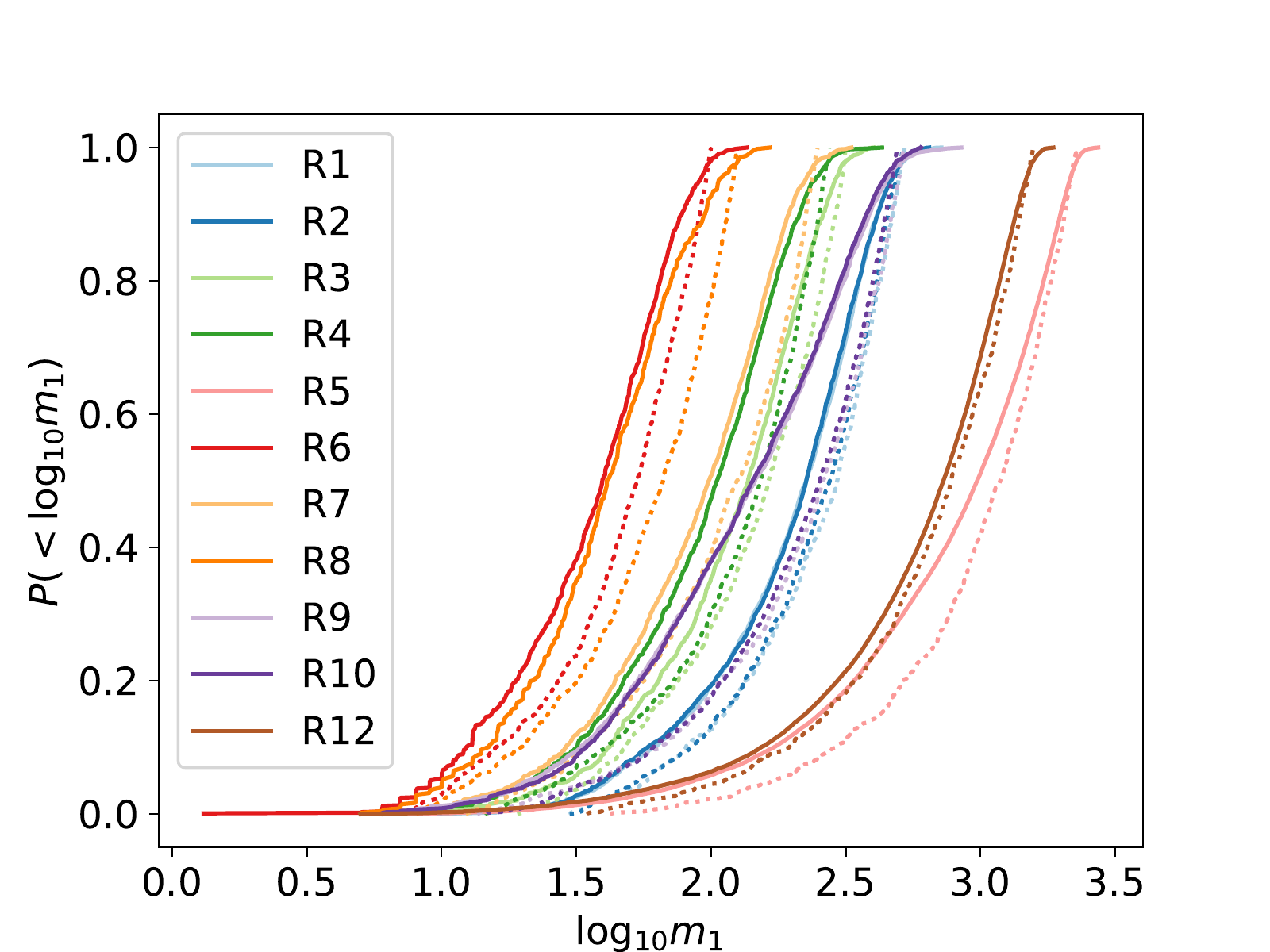}
\includegraphics[width=\columnwidth]{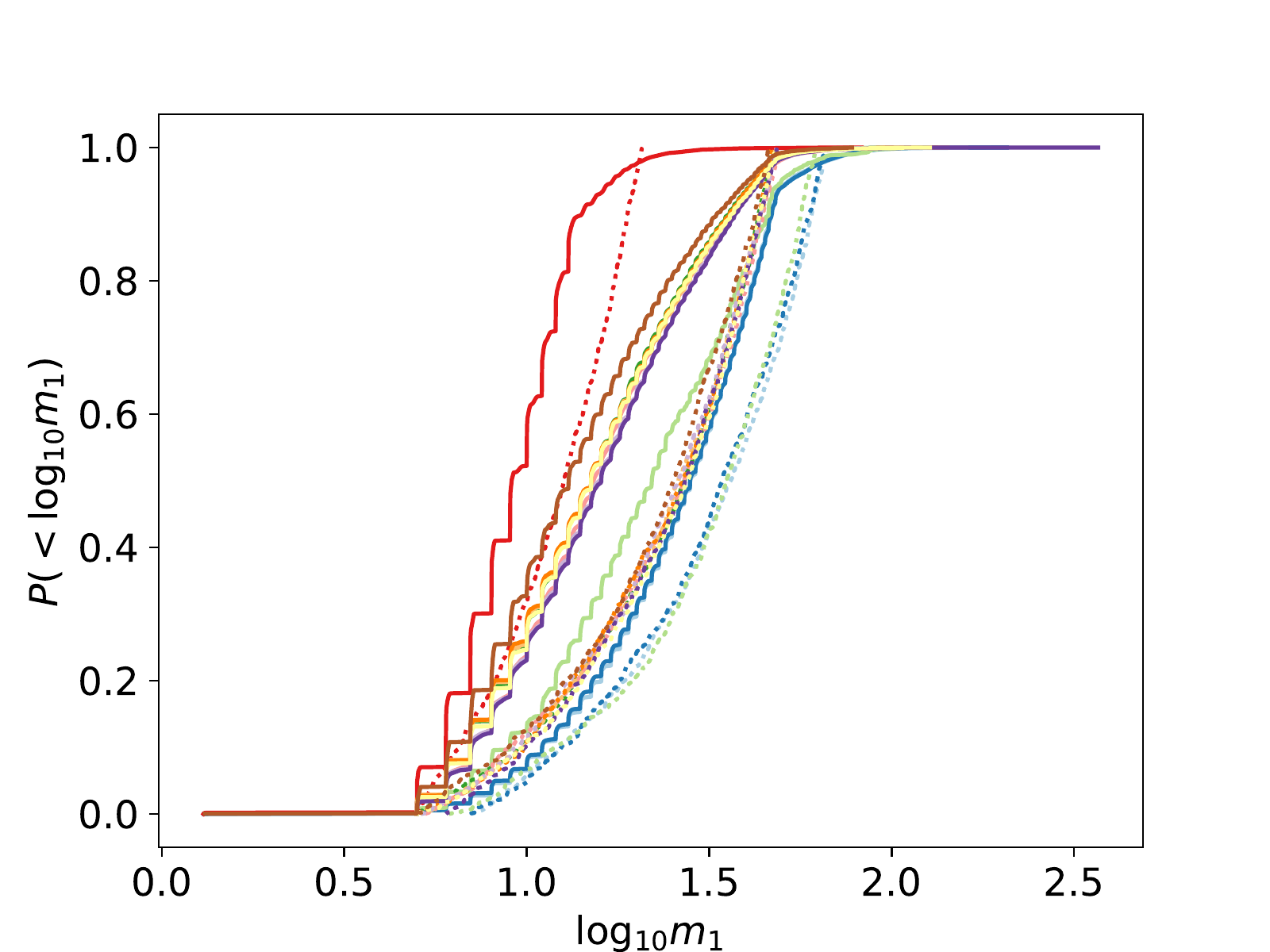}
\caption{\label{fig:PrimaryMass:Empirically} 
\emph{Top panel} (trap) shows the CDF for $m_1$ in the trap.  Dashed lines show an approximately uniform distribution
$P(<m_1)\propto (m_1-m_{\mathrm{min}})/(m_{\mathrm{max}}-m_{\mathrm{min}})$ for suitable $m_{\mathrm{max}},m_{\mathrm{min}}$.
\emph{Bottom panel} (bulk) shows the CDF for $m_1$ in the bullk.  Dashed lines show an approximation of a uniform
distribution in $\log m_1$ (i.e., as if $\gamma=1$). 
}
\end{figure}

\begin{figure}
% python3 plot_distribution_and_model.py --directory ../data/simulations/v3f/ trap  m_2 --x-log-scale --showdist 1
% python3 plot_distribution_and_model.py --directory ../data/simulations/v3f/ bulk  m_2 --x-log-scale --showdist 2
\includegraphics[width=\columnwidth]{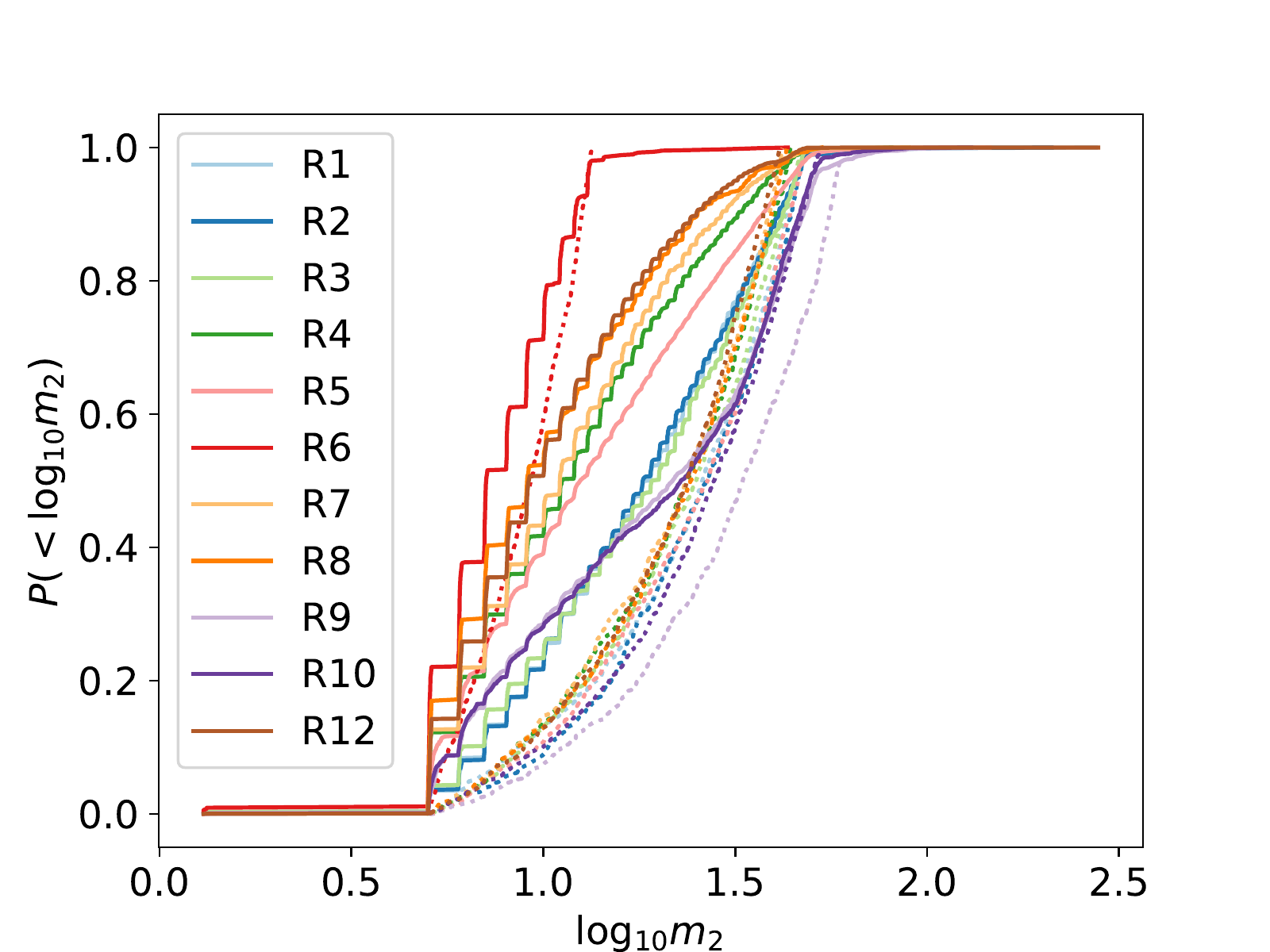}
\includegraphics[width=\columnwidth]{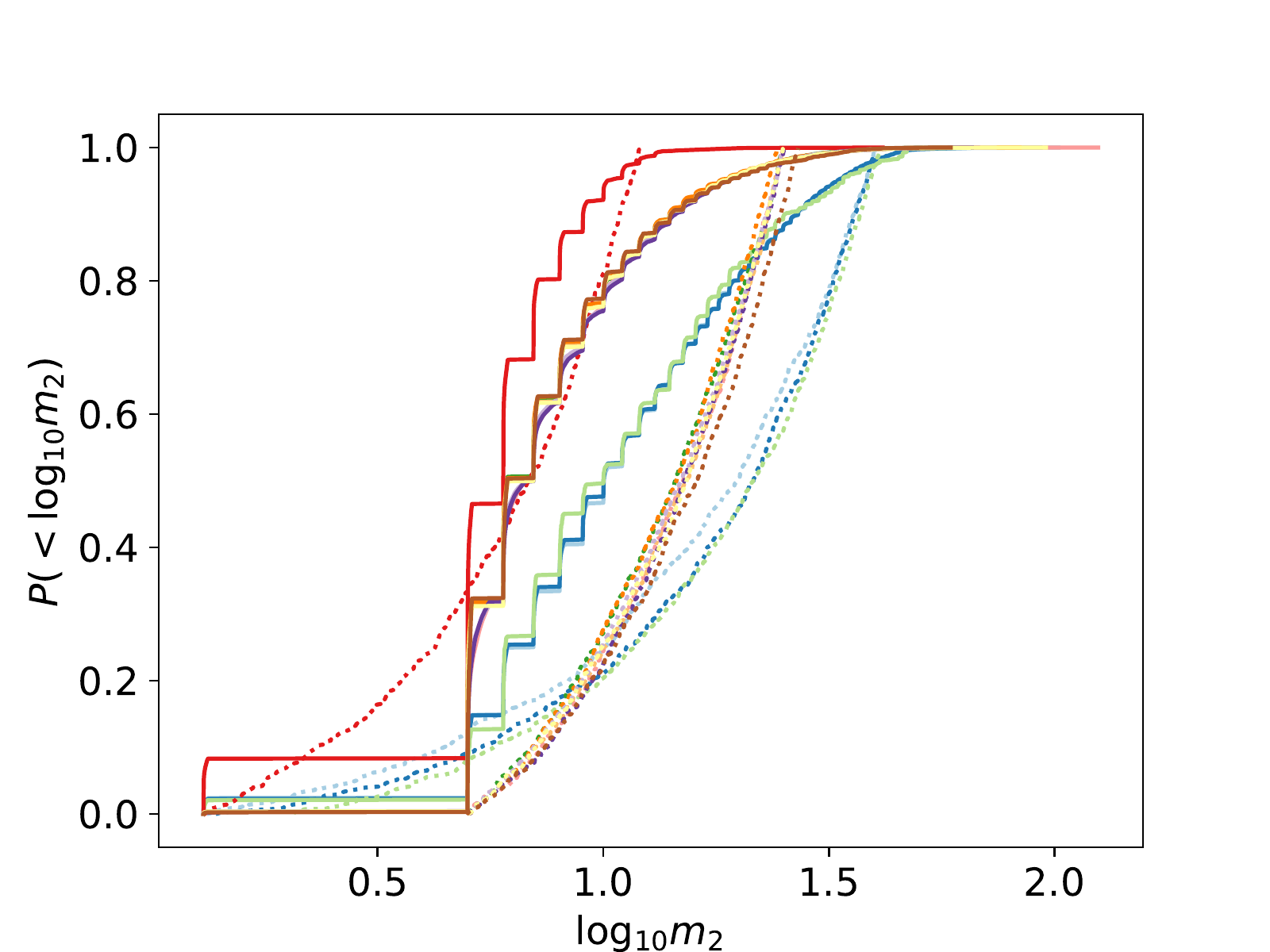}
\caption{\label{fig:SecondaryMass:Empirically} 
\emph{Top panel} (trap) shows the CDF for $m_2$ in the trap.  Dashed lines show an approximation with $\gamma_{eff}=1$
\emph{Bottom panel} (bulk) shows the CDF for $m_2$ in the bulk. Dashed lines show an approximation with $\gamma_{\rm
  eff}=4$.
Note that in all cases, $m_2$ is almost always bounded by the maximum mass in the progenitor population -- in other
words, essentially no second-generation objects are secondary masses $m_2$
}
\end{figure}

Figures \ref{fig:PrimaryMass:Empirically} and \ref{fig:SecondaryMass:Empirically} motivate our phenomenology.   These figures show the (cumulative) distribution of masses $m_1$ and $m_2$ due to mergers in the bulk and in the trap, averaged over time.    All mass distributions are well-approximated by simple power laws.  
In the trap, the distribution of primary masses $m_1$ is dominated by IMBHs, which grow linearly with time in the trap up to a large (and physics-dependent) cutoff and which therefore have a uniform time-averaged mass distribution.  The secondary mass distribution, by contrast, is a power law bounded by the same mass limits as the input BH population.  The relatively stable upper mass limit is striking and important: despite some growth, and despite mechanisms that modify the power law to more strongly favor higher-mass secondaries $m_2$ in the trap, the secondary masses in BBH trap mergers do not significantly extend to higher masses.
In the bulk, the distribution of primary and secondary masses are dominated by the input population; for example, when the input mass distribution has $\gamma=2$, the primary mass distribution also a power law $\gamma_{\rm eff} \simeq 2$.   Migration modifies the mass distribution for the secondary population, which has an effective power law $\gamma_{\rm eff}\simeq 4$.

This phenomenology does not  illustrate a notable signature of 1g or 2g components assembled in the bulk of the disk.  While our previous calculations demonstrate such mergers occur, when viewed via GW alone  averaged over time and without a priori knowledge of the IMF in the disk, we do not have a compelling \emph{phenomenological} reason to include them as a seperate population.  Instead, for the immediate future and given the sparsity of the current observational sample, we will model the joint bulk and 1g/2g population into a single component, with a common power law with variable exponent and upper limit.


Motivated by our calculations, we adopt a simple two-component mass model, superposing phenomenological approximations to the distribution of ``trap'' mergers and ``bulk'' mergers.  We treat these components as two completely independent power-law distributions, not imposing any consistency requirements on the relative rates or power laws aside from the requirement that they share a common maximum mass $m_{bh,max}$ for the progenitor population.  Specifically, we use the following merger rate model:

\begin{equation}
\rho(m_1,m_2) = \frac{dN}{dVdt} =
\rho_{\mathrm{bulk}} + \rho_{\mathrm{trap}}
\end{equation}
\begin{align}
\rho_{\mathrm{bulk}} &=
\mathcal{R}_{\mathrm{bulk}} \,
\mathcal{P}(m_1 | \alpha_{\mathrm{bulk}}, m_{\mathrm{min}}, m_{\mathrm{max,\star}}) \,
\mathcal{P}(m_2 | \alpha_{\mathrm{bulk}}, m_{\mathrm{min}}, m_{\mathrm{max,\star}})
\\
\rho_{\mathrm{trap}} &=
\mathcal{R}_{\mathrm{trap}} \,
\mathcal{U}(m_1 | m_{\mathrm{min}}, m_{\mathrm{max,IMBH}}) \,
\mathcal{P}(m_2 | \alpha_{\mathrm{trap}}, m_{\mathrm{min}}, m_{\mathrm{max,\star}})
\end{align}
where $\mathcal{U}(x | x_{\mathrm{min}}, x_{\mathrm{max}})$ denotes a uniform distribution in $x$, on the interval $[x_{\mathrm{min}}, x_{\mathrm{max}}]$, and $\mathcal{P}(x | \alpha, x_{\mathrm{min}}, x_{\mathrm{max}})$ denotes a powerlaw distribution proportional to $x^{-\alpha}$, on the interval $[x_{\mathrm{min}}, x_{\mathrm{max}}]$.  In all runs, we set the minimum black hole mass to $m_{\mathrm{min}} = 5M_\odot$ and allow the maximum IMBH mass $m_{\mathrm{max,IMBH}}$ to vary freely, with a uniform-in-log prior.  We then have one run with the stellar-mass black hole mass fixed to $m_{\mathrm{max,\star}} = 50 M_\odot$, to match our simulations, as well as an additional run where we allow it to vary freely, to account for uncertainties in our simulations, with a uniform-in-log prior.  Notably, the ``bulk'' component of this model closely resembles power-law mass distributions previously used to interpret GW source populations; see, e.g., \cite{Abbott16b,Wysocki18}.
For simplicity, because our BH spins distributions are broad and don't show strong imprints of hierarchical growth, we adopt a fiducial spin distribution in our analysis and recovery, effectively ignoring the prospects for using spin to distinguish between different formation options. 
%\ros{will add spin, but it is not active as of today, and will be a common model}

We compare this phenomenological model to synthetic and real GW observations using the \textsc{PopModels}  code from \cite{Wysocki18}.   This code infers source populations, given observations and a model for detector sensitivity.  Unless otherwise noted, we will use a simplified model for detector sensitivity that accounts for BH mass and (nonprecessing) spin, based on a single interferometer's sensitivity to nonprecessing black holes and assuming an analysis minimum frequency of $f_{\mathrm{min}}=10\,\mathrm{Hz}$.
%\dwy{Note we evolved the waveforms from 1Hz, but performed the integral over the PSD from 10Hz}.

\begin{figure}
\includegraphics[width=\columnwidth]{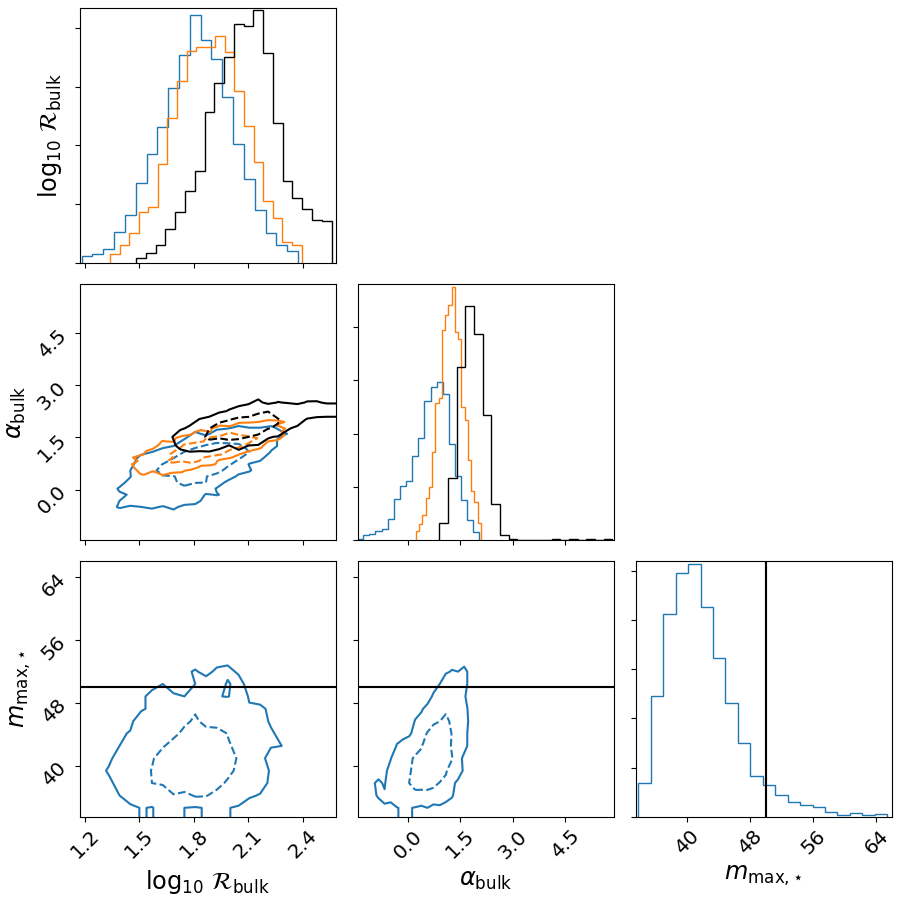}
\includegraphics[width=\columnwidth]{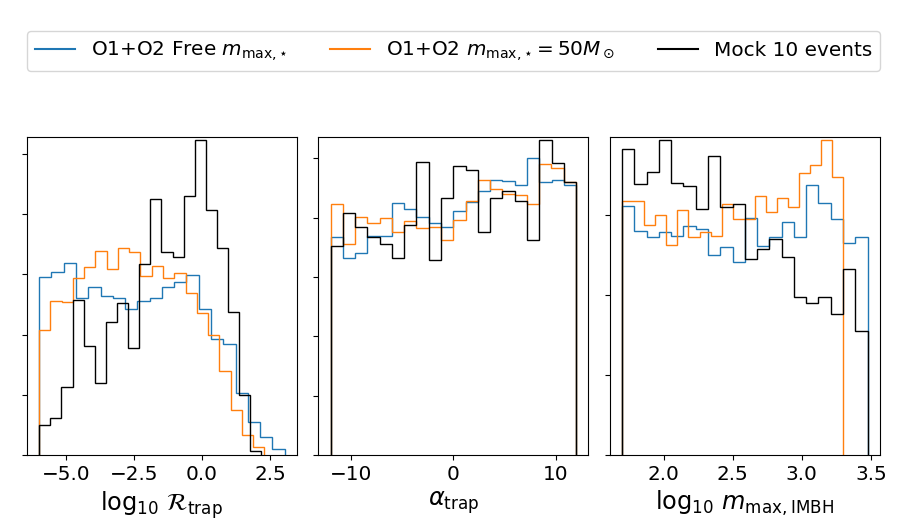}
\caption{\label{fig:PhenomenologyResultsO2}Inferences about a phenomenological AGN disk population, using GW observations available in \citep{Abbott18Pop}, as well as mock data from a representative realization of this model.  Each panel shows either one- or two-dimensional marginal distributions for our models' phenomenological parameters: the merger rates $R_{\rm{bulk}},R_{\rm{trap}}$; the power law index $\alpha$ in the bulk and trap; and the upper mass limits $m_{\mathrm{max}}$ for stellar- and intermediate-mass BHs. See Table \ref{tab:O2LIGO} for credible intervals for these parameters. }
\end{figure}

A direct observational constraint on AGN disk BBH formation, Figure \ref{fig:PhenomenologyResultsO2} shows our inferences about phenomenological AGN disk population parameters given current reported GW observations to date.   In the absence of reported IMBH observations which could distinctively identify a trap component, our conclusion is an upper limit on the AGN disk component of 161 Gpc$^{-3}$ yr$^{-1}$ from the trap (alone) and $143$ Gpc$^{-3}$ yr$^{-1}$ from the bulk disk (alone); see Table \ref{tab:O2LIGO}.

\begin{table*}
 \begin{minipage}{140mm}
\caption{ \label{tab:O2LIGO} Inferences about our AGN disk phenomenological model, given reported observations in \citep{Abbott18Pop} and illustrated in Figure~\protect{\ref{fig:PhenomenologyResultsO2}}. Limits are expressed as 90$\%$ credible intervals on each population hyperparameter, except for $m_{\rm{max,IMBH}}$ which is a $95\%$ lower limit.  All rates are in $\rm{Gpc}^{-3} \rm{yr}^{-1}$, and all masses in $M_{\odot}$.}
    \begin{tabular}{l|cc|cc|cc|cc|cc|c}
   Model & $\mathcal{R}_{\mathrm{trap}}$ & & $\alpha_{\mathrm{trap}}$ & & $\mathcal{R}_{\mathrm{bulk}}$ & & $\alpha_{\mathrm{bulk}}$ & & $m_{\mathrm{max},{\star}}$ & & $m_{\mathrm{max,IMBH}}$ \\ \hline
   Fixed&  $0.00$ & $4.95$ & $-10.82$ & $10.90$ & $34.91$ & $165.71$ & $0.61$ & $1.81$ & --- & --- & 60.17 \\
   Free & $0.00$ & $14.02$ & $-10.64$ & $10.92$ & $28.77$ & $143.09$ & $-0.40$ & $1.56$ & 36.07 & 51.39 & 60.01\\
    \end{tabular}
    \end{minipage}
\end{table*}

\begin{figure}
\includegraphics[width=\columnwidth]{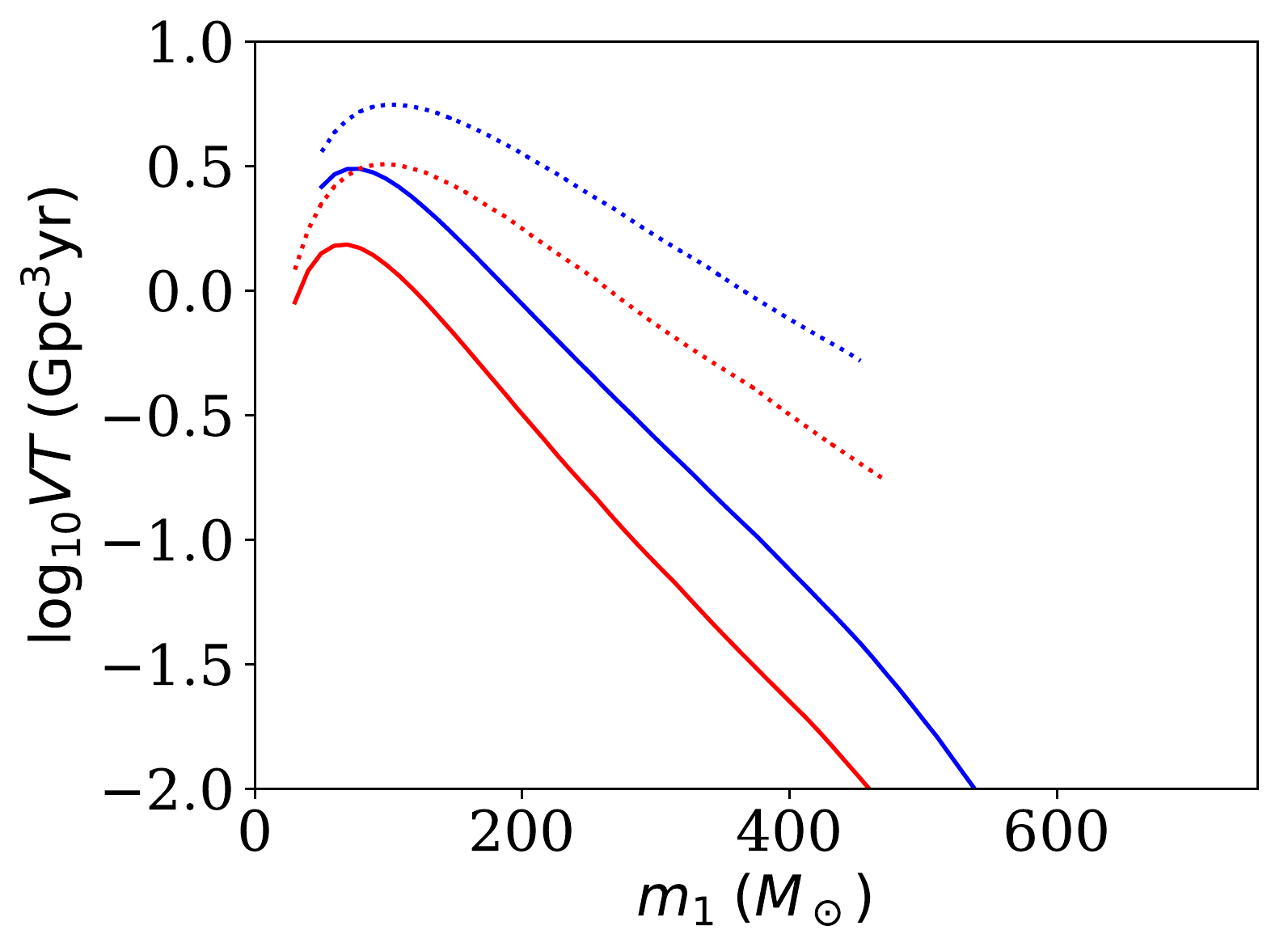}
\includegraphics[width=\columnwidth]{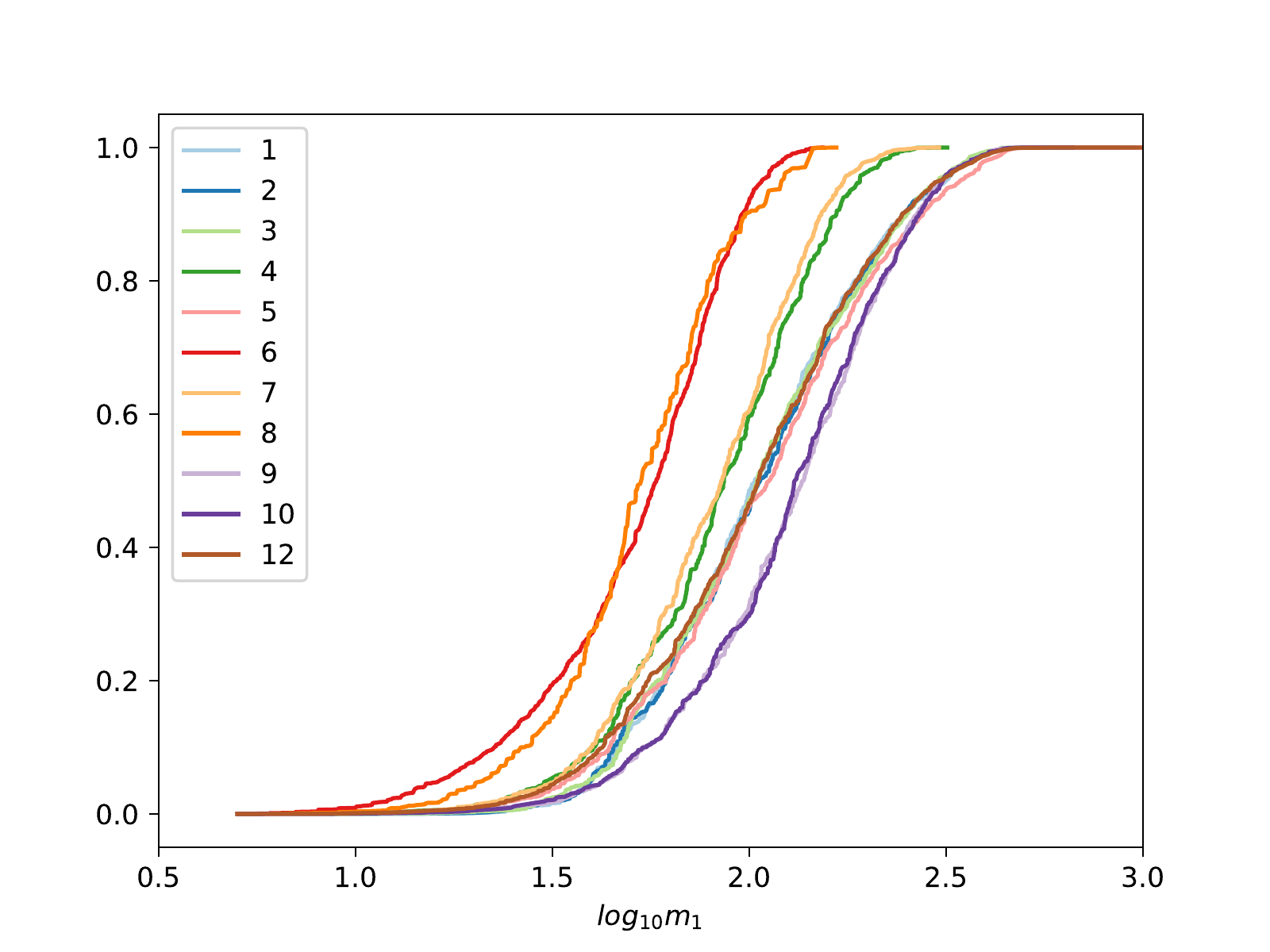}
\caption{\label{fig:Sensitivity}\emph{Top panel}: Plot of network sensitivity $VT$ versus $m_1$, assuming $m_2=50 M_\odot$ (blue) or $30 M_\odot$ (red), and a low-frequency analysis cutoff of $f_{\rm min}=10{\rm Hz}$.  Solid lines show zero BH spin; dotted lines assume the IMBH has spin $a_1=0.8$ and the merger occurs in a prograde orbit.  Due to limited low-frequency sensitivity, GW detectors may have strongly suppressed sensitivity to the most high-mass and asymmetric sources produced in AGN migration traps, unless the IMBHs are spinning rapidly.
\emph{Bottom panel}: %\dwy{Needs y-axis label}
Detection-weighted cumulative distribution of $m_1$ for BBH mergers in the trap, accounting for BH spins, to contrast to Figure \ref{fig:PrimaryMass:Empirically}.  Due to limited sensitivity to $m_1>300 M_\odot$ indicated in the top panel, the IMBH-BH mergers observed in GW in the near future will not always probe the most massive IMBHs produced in AGN disks, without lower-frequency analysis.
}
\end{figure}

To assess the prospects for constraining AGN disk formation in the near future, we generate several synthetic binary black hole populations drawn from our simulations and from our phenomenological models, and apply the same procedure.   To better understand our results, the top panel of Figure \ref{fig:Sensitivity} shows  the interferometer network sensitive volume as a function of total mass for binaries with $m_2=50 M_\odot$ versus $m_1$, for a few choices of BH spin.  While network sensitivity increases rapidly at low mass, eventually as the primary black hole mass grows larger, the associated GW merger signal is produced at or below the minimum sensitive frequency viable for analysis at present.  
As shown in the bottom panel of  Figure \ref{fig:Sensitivity}, the strong mass dependence of network sensitivity influences the  \emph{detection-weighted} mass distribution (i.e., the mass distribution of observed BHs), so much so that the largest IMBH binaries produced in each population are  not reliably observed.   In other words, because of the low-frequency sensitivity of currently-operating GW detector networks, the most massive IMBHs $m_1> 300M_\odot$ by this channel are hard to find when paired with stellar-mass companions as predicted here.  For this reason, constraints on the maximum IMBH mass are expected to be weak, unless that mass limit falls below $300 M_\odot$.

Within this phenomenological approach, only by finding IMBHs can we identify the unique signature of the trap component and thus measure ${\cal R}_{\rm trap}$.  As illustrated by the detection volume and detection-weighted mass distributions in Figure \ref{fig:Sensitivity}, however, ground-based sensitivity to very massive IMBHs \emph{in asymmetric and often high mass ratio binaries} is relatively small.  The ratio of detected binaries in the trap versus the bulk is proportional to their detection-volume-weighted merger rates, with a ratio  $F \simeq V(m_2\simeq 50, m_1){\cal R}_{\rm trap}]/[V(m_1,m_2\simeq 30 M_\odot) {\cal R}_{\rm bulk} $), where $m_1\simeq 200 M_\odot$ is a typical detectable IMBH mass.   The ratio will be further reduced by approximately a factor ${\cal G}=300M_\odot/m_{IMBH,max}$, to account for the nondetectability of very massive high-$q$ IMBH binaries; for our reference population, with $m_{IMBH,max}=2500$, this factor is ${\cal G}\simeq 0.1$.   If GW observatories continue to exclude IMBHs, then ${\cal R}_{\rm trap}/{\cal R}_{\rm bulk}$ is constrained by $F\lesssim 1/N_{obs}{\cal G}$.
To illustrate this plausible scenario, we generated a synthetic population with ${\cal R}_{bulk}=150 {\rm Gpc}^{-3}{\rm yr}^{-1}$ and ${\cal R}_{\rm trap}=30 {\rm Gpc}^{-3}{\rm yr}^{-1}$, with the mass distribution of the bulk BBH population drawn from previously reported inferences about observational results.    For simplicity and to be pessimistic, we performed our analysis assuming BH spins do not impact detector network sensitivity.   In this synthetic scenario, no binaries with an IMBH are found within the first 95 observations, as they make up $\mathcal{O}(1\%)$ of detections under this model. Therefore our analyses of these events (or any smaller subset) only recover the properties of the injected bulk population,  similar to Figure \ref{fig:PhenomenologyResultsO2}.
Because many of the synthetic trap IMBH binaries are very massive, our constraints on the overall merger rate are weak.

%\ros{finish: synthetic recovery} 
 
\section{Conclusions}
As LIGO identifies the distribution of BHs involved in mergers in our Universe, we gain increasingly strong astrophysical constraints on AGN disks. The merger of BHs in AGN disks directly probes physical parameters like gas density, gas distribution and disk lifetime, which are hard to extract from electromagnetic observations. For example, LIGO is presently telling us that low ionization nuclear emission regions (LINERs) cannot be mostly optically-thick advection dominated accretion flows \citep{McK17,Ford19}.  Future LIGO population statistics will constrain these astrophysical parameters directly and LIGO's results over the next decade will have broad astrophysical implications for models of $\Lambda$ CDM feedback, AGN formation, fuelling and extinction.\\

In this work we provide the mass and $\chi_{\rm eff}$ distributions under various assumptions about the physics involved in this compact binary formation channel. 
% Mass distributions
For the bulk of the disk, the mass ratio distribution among merging binaries is centered around $\tilde{q} \sim 0.5-0.7$.  Even if we adopt  a black hole IMF consistent with Galactic BH masses ($5-15M_{\odot}$), we also find a modestly asymmetric population $\tilde{q} \sim 0.7$.   
The migration trap will steadily grow an IMBH,  whose maximum mass reflects the disk lifetime and natal BH population.   Mergers of BH binaries in that trap are much more asymmetric than mergers in the bulk of the disk, with typical mass ratios $\tilde{q}\sim 0.1$.
Thus, ongoing observational constraints on the high-mass  BH population from LIGO put limits on AGN disk lifetime and disk structure (in particular the existence of migration traps) as well as orbital grind-down efficiency.
Conversely, ongoing LIGO non-detections of black holes $>10^{2}M_{\odot}$ puts strong limits on the presence of migration traps in AGN disks and the AGN disk lifetime. 

% Spin distribution
The AGN disk channel can impart a distinctive spin distribution to the IMBH population.  That said, for the majority of mergers, the spin  distribution reflects assumptions about the natal spin distribution, convolved with the spin alignment physics discussed in the text.  In our calculations, the $\chi_{\rm eff}$ distribution from BH mergers in the bulk of the disk is naturally centered around ($\tilde{\chi}_{\rm eff} \approx 0$).   The median of $\tilde{\chi}_{\rm eff}$ is slightly negative if gas hardening of retrograde binaries is more efficient than for prograde binaries, while  the width of the $\chi_{\rm eff}$ distribution is significantly narrower for narrow spin magnitude distributions.      Inverting from the observed population as LIGO continues to operate, an observed bias towards small positive or negative $\tilde{\chi}_{\rm eff}$ may allow us to distinguish the efficiency of gas hardening of prograde or retrograde binaries. 

% Other

The rate of black hole mergers is highest early on ($\sim 0.1$Myr) in the AGN disk lifetime. The highest merger rate occurs for this channel if AGN disks are relatively short-lived ($\sim 1$Myr) so multiple AGN episodes can happen per Galactic nucleus in a Hubble time. Finally, we wish to emphasize that LIGO observations going forward will allow us to dramatically constrain models of AGN disks in the next few years.

{\section{Acknowledgements.}}BM \& KESF are supported by NSF grant 1831412. ROS and DW are supported by NSF PHY-1707965; DW also thanks the RIT COS and CCRG for support. We acknowledge very useful discussions with Bernard Kelly, Cole Miller, Yuri Levin and Phil Armitage.

\end{document}